%% file: AGRXAI.tex
\documentclass[lettersize,journal]{IEEEtran}
\usepackage{amsmath,amsfonts}
\usepackage{algorithm}
\usepackage{array}
\usepackage[caption=false,font=normalsize,labelfont=sf,textfont=sf]{subfig}
\usepackage{textcomp}
\usepackage{stfloats}

\usepackage{url}
\usepackage{verbatim}
\usepackage{graphicx}
\usepackage{cite}
\usepackage{algpseudocode}
\usepackage{multirow}
\usepackage{shortcuts}
\usepackage{tabularx}
\usepackage{booktabs}
\usepackage{subfig}
\usepackage{caption}

\let\oldsubfloat\subfloat
\renewcommand\subfloat[2][\relax]{\oldsubfloat[\footnotesize#1]{#2}}

\usepackage[flushleft]{threeparttable}
\begin{document}

\title{\codename: Defending Federated Learning Against Poisoning Attacks Through Local Update Amplification}

\author{Zirui Gong$^*$, \and 
	Liyue Shen$^*$, \and
	Yanjun Zhang, \and
	Leo Yu Zhang,  \and
	Jingwei Wang,  \and
	Guangdong Bai, \and
	and Yong Xiang 

\IEEEcompsocitemizethanks{
\IEEEcompsocthanksitem This article extends the preliminary results presented in~\cite{shen_better_2022}. In our prior work, we focus on a basic approach \agrmp, which draws inspiration from the max pooling operation to amplify the salient
features of local updates. In this work, we introduce a more advanced approach named \agrxai to enhance the distinctions between benign gradients and malicious gradients. The paper also substantially extends experimental evaluation on seven datasets across diverse domains, demonstrating the consistent improvement of robustness and fidelity.
\IEEEcompsocthanksitem Correspondence to Dr. Y. Zhang and Dr. L. Zhang.
\IEEEcompsocthanksitem $^*$ These authors contributed equally to this work.
\IEEEcompsocthanksitem Zirui Gong and Leo Yu Zhang are with the School of Information and Communication Technology,  Griffith University, Southport, Queensland, Australia. Email: z.gong@griffith.edu.au, leo.zhang@griffith.edu.au.
\IEEEcompsocthanksitem Liyue Shen, Jingwei Wang, and Guangdong Bai are with the School of Information Technology and Electrical Engineering, The University of Queensland, St Lucia, Queensland, Australia. Email: liyue.shen@uq.net.au, jingwei.wang@uq.net.au, g.bai@uq.edu.au.
\IEEEcompsocthanksitem Yanjun Zhang is with the School of Computer Science, University of Technology Sydney, Sydney, New South Wales, Australia. Email: Yanjun.Zhang@uts.edu.au. 
\IEEEcompsocthanksitem Yong Xiang is with the School of Information Technology, Deakin University, Melbourne, Victoria, Australia. E-mail: yong.xiang@deakin.edu.au.
}

}

\maketitle
\input{chapters/abstract}
\begin{IEEEkeywords}
Federated Learning, Byzantine-robust Aggregation, Poisoning Attack, Explainable AI.
\end{IEEEkeywords}

\input{chapters/intro1}

\input{chapters/background1}

\input{chapters/method}
\input{chapters/setup}
\input{chapters/experiment}
\input{chapters/discussion}

\input{chapters/conclusion}



\bibliographystyle{IEEEtran}
\bibliography{references}

\input{chapters/Appendix}

\end{document}

%% file: chapters/abstract.tex
\begin{abstract}
The collaborative nature of federated learning (FL) poses a major threat in the form of manipulation of local training data and local updates, known as the Byzantine poisoning attack. 
To address this issue, many Byzantine-robust aggregation rules (\agrs) have been proposed to filter out or moderate suspicious local updates uploaded by Byzantine participants.  

This paper introduces a novel approach called \codename, aiming to simultaneously improve robustness, fidelity, and efficiency of the existing \agrs. 
The core idea of \codename is to amplify the ``morality'' of local updates by identifying the most repressive features of each gradient update, which provides a clearer distinction between malicious and benign updates, consequently improving the detection effect. 
To achieve this objective, two approaches, namely \agrmp and \agrxai, are proposed. \agrmp organizes local updates into patches and extracts the largest value from each patch, while \agrxai leverages explainable AI methods to extract the gradient of the most activated features. By equipping \codename with the existing Byzantine-robust mechanisms, we successfully enhance the model robustness, maintaining its fidelity and improving overall efficiency.

\codename is universally compatible with the existing Byzantine-robust mechanisms. The paper demonstrates its effectiveness by integrating it with all mainstream \agr mechanisms. 
Extensive evaluations conducted on seven datasets from diverse domains against seven representative poisoning attacks consistently show enhancements in robustness, fidelity, and efficiency, with average gains of 40.08\%, 39.18\%, and 10.68\%, respectively.

\end{abstract}

%% file: chapters/intro1.tex
\section{Introduction}

\IEEEPARstart{F}{ederated} learning (FL) \cite{mcmahan_communication-efficient_2017,flSurvey,zhu2021semi,zheng2022aggregation,zeng2023hfedms} is a subset of the deep learning system, where several clients collaboratively train a central model. It has become more prevalent across a range of privacy-sensitive tasks and has been implemented by well-known machine learning techniques. In FL, each client maintains a private training dataset and trains a local model based on their datasets. After each local training, clients only update the gradient to the central server (i.e., the aggregator) for the global model aggregation. On the server side, the global model is updated by taking one step downward in gradient descent; then, the global model is further broadcasted to clients for the subsequent training cycle. FL enables the training of high-quality Machine Learning (ML) models with massive data and eliminates the exposition of private raw data to the server. 

However, the distributed nature of FL makes it susceptible to client-side poisoning attacks~\cite{bhagoji_analyzing_2019,bagdasaryan_how_2019, cao_fltrust_2022,fang_local_2021,kurakin_adversarial_2017,li_lomar_2022,munoz-gonzalez_towards_2017,noauthor_manipulating_nodate, zhang2023agrevader, wei2023client}. 
One kind called untargeted attacks~\cite{cao_fltrust_2022,fang_local_2021,kurakin_adversarial_2017,li_lomar_2022,munoz-gonzalez_towards_2017,noauthor_manipulating_nodate}, which aims to corrupt the global model to low test accuracy, therefore causes the model unusable and eventually leads to denial-of-service attacks (e.g., an attacker may perform such attacks on its competitor’s FL system). On the other hand, targeted attacks~\cite{bhagoji_analyzing_2019,bagdasaryan_how_2019,wang2021neural}, also known as the backdoor attack, where the attacker corrupts the global model to predict an attacker-chosen label for any testing input embedded with a trigger while maintaining high test accuracy on other input.


Several mechanisms have been proposed to defend against the poisoning attacks \cite{zhu2023adfl,xie2021crfl,yin_byzantine-robust_2018,cao_fltrust_2022,fang_local_2021,mhamdi_hidden_2018,li_lomar_2022,shen_better_2022,zhang2022fldetector,geng2023better,xu2022communication, ma2023loden}.
One representative method is called Byzantine-robust defense mechanisms 
~\cite{yin_byzantine-robust_2018,cao_fltrust_2022,fang_local_2021,mhamdi_hidden_2018,li_lomar_2022,shen_better_2022,zhang2022fldetector,geng2023better,xu2022communication}, in which the server employs a robust aggregation algorithm (\agr) to filter out or moderate suspicious local updates to mitigate the malicious impact of adversary contributions. There are three widely used \agr mechanisms, i.e., distance-based
\cite{yin_byzantine-robust_2018,mhamdi_hidden_2018}, prediction-based \cite{fang_local_2021,zhang2022fldetector,li_lomar_2022}, and trust bootstrapping-based techniques \cite{cao_fltrust_2022,geng2023better,xu2022communication}. Specifically, the distance-based mechanism compares the collected gradients based on distance measurements, i.e., Euclidean distance and cosine similarity, and removes anomalous updates before aggregating the remaining ones. The prediction-based mechanism checks the model's prediction performance and removes the updates causing performance degradation, while the trust bootstrapping-based mechanism computes trust scores for each participant and uses them as weights when averaging updates. 


\textbf{Triad of Byzantine-robust FL:} 
We articulate the triad of desirable properties of \agrs as follows. 
(1) \textbf{\emph{Robustness}}. The \agrs shall minimize the decrease of the global model’s test accuracy caused by malicious updates. (2) \textbf{\emph{Fidelity}}. 
The \agrs shall not harm the performance of the global model when there are no malicious updates and shall achieve a test accuracy close to that of the plain FedAvg~\cite{cao_fltrust_2022}.  
Additionally, 
the \agrs should be designed to handle the inherent diversity of data from different owners. While this data variety can enhance the generalization of the global model, it can also result in deviations in the local updates, which existing defenses may deny. 
(3) \textbf{\emph{Efficiency}}. 
The method shall retain the computation efficiency and scale in large-scale FL training, particularly in processing the high-dimensional local updates from a large number of participants. 
The primary question this paper addresses is how to achieve and enhance the triad of robustness, fidelity, and efficiency for Byzantine-robust aggregators in FL.


\textbf{Our work:} 
We introduce a novel approach called \codename, which is built upon the base \agrs, aiming to achieve robustness, fidelity, and efficiency in FL.
The central idea behind \codename is to amplify the influence of both malicious and benign local updates by focusing on the gradients of the most prominent features. Through this, we can make better decisions about removing malicious updates before global model aggregation.
To achieve this goal, we develop two strategies, \agrmp and \agrxai.

The \agrmp draws inspiration from the widely used max pooling operation \cite{yamaguchi_neural_1990} in  Convolutional Neural Networks (CNN), which can reduce the dimensionality of features and summarize their most activated presence. By leveraging this technique, \agrmp seeks to improve the aggregation process by amplifying the salient features of local updates, thereby enhancing their impact on the final aggregated output. Specifically, \agrmp initially rearranges each updated gradient into individual patches. Subsequently, it identifies the most prominent feature in each patch by extracting the largest value and returns it as the amplified outcome. 

The \agrxai is motivated by recent progress in the integration of XAI within the cybersecurity domain \cite{charmet2022explainable,kuppa2021adversarial} and applies the concept of Explainable AI (XAI) \cite{linardatos_explainable_2020} to amplify the difference between benign gradients and malicious gradients. 
Specifically, we employ Grad-CAM \cite{selvaraju_grad-cam_2020} to determine the feature maps that learned the key features.  
After getting the importance weights, we rank them in descending order and select the top $p$ feature maps with the highest weights ($p$ indicates the percentage we extracted). This selection allows us to focus on the most significant feature maps. By indexing the gradients corresponding to these selected feature maps, we can extract the top $p$ most significant gradients as the amplified outcome.  

\begin{figure}
\vspace{-\baselineskip}
\centering
    \subfloat[Untargeted attack]{\includegraphics[width=1.8in]{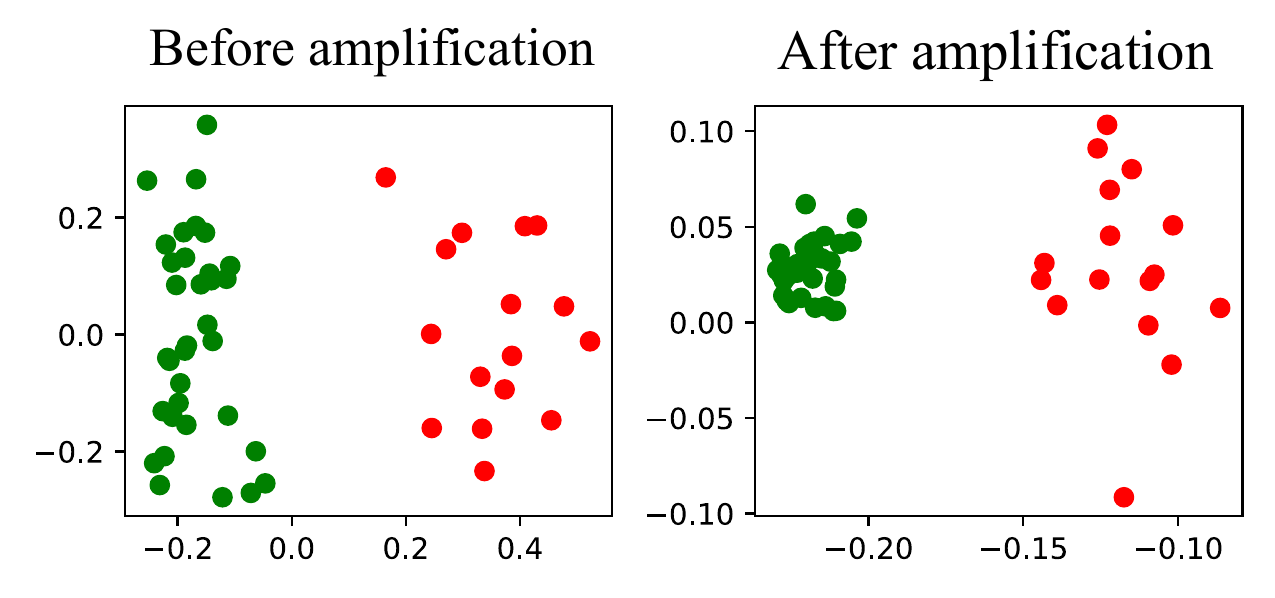}%
\label{fig:pca-untargetd}}
%
\subfloat[Targeted attack]{\includegraphics[width=1.8in]{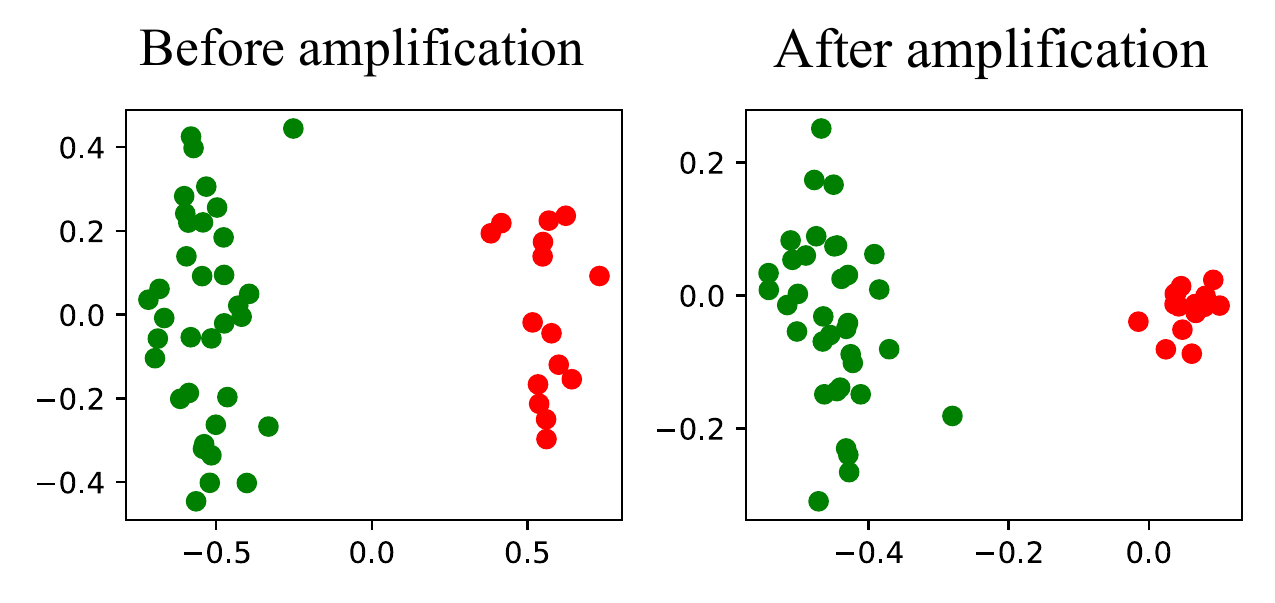}%
\label{fig:pca-target}}
\caption{Utilizing the PCA method to project gradients onto a two-dimensional surface. Specifically, we plot a total of 50 local updates at the 70th epoch of the training process using the LOCATION30 dataset \cite{yang_participatory_2016}. Within the plotted updates, red dots represent malicious updates, while green dots represent benign ones. The attack employs the untargeted label flip \cite{bagdasaryan_how_2019} and targeted scale attack \cite{cao_fltrust_2022}.} 
\label{fig:pca}
\end{figure}

The seemingly straightforward strategies are effective in achieving the triad of Byzantine-robust FL. 
Firstly, the extraction of the most activated gradients renders the local updates more distinguishable after the amplification. As illustrated in Fig.~\ref{fig:pca}, the benign gradient is reduced from [-0.3, 0.4] to [0, 0.06] (the y-axis of green dots in Fig. \ref{fig:pca-untargetd}), while the malicious gradient is reduced from [-0.4, 0.3] to [-0.1, 0.1] (the y-axis of red dots in Fig. \ref{fig:pca-target}).
Therefore, the \agrs can make robust decisions of detecting maliciousness local updates (detailed in Section \ref{sec:robust}). 
Secondly, \codename enhances fidelity by providing invariance to distortion arising from local translations. Analogous to a max pooling layer in typical CNNs, \codename can suppress small changes. Hence, when no attack is present, it functions as a noise canceler (detailed in Section \ref{sec:Fidelity}). 
Thirdly, the significant dimension reduction in the feature space provides substantial benefits to the efficiency of \agrs. As many existing \texttt{AGR}s \cite{cao_fltrust_2022,fang_local_2021} exhibit superlinear time complexity and \codename amplifies the advantages derived from the input size reduction (detailed in Section \ref{sec:Efficiency}).

It is important to note that the proposed amplification process is universally compatible with any existing \agrs, regardless of the underlying aggregation rules. In this study, we applied \codename to three widely used mechanisms, including distance-based, prediction-based, and trust bootstrapping-based aggregators. Specifically, we introduce ten variations of \codename, i.e., \cosmp, \emp, \mmp, \fangmp, \flmp, \cosxai, \exai, \mxai, \fangxai, and \flxai. Each of these variations is suitable for different use cases, which are explored and discussed in Section \ref{sec:compare}.
Overall, our results consistently demonstrate improvement for all three mechanisms. For instance, when the distance-based mechanisms are equipped with \codename, they outperform their base versions by 66.26\% in robustness, 29.6\% in fidelity, and 12.9\% in efficiency. Similarly, in the case of prediction-based mechanisms, \codename yields 35.59\% improvement in robustness, 47.3\% improvement in fidelity, and 7.7\% improvement in efficiency compared to its counterpart. Furthermore, \codename also enhances the performance of the trust bootstrapping-based mechanisms in the state-of-the-art aggregator by 18.37\% in robustness, 19.4\% in fidelity, and 12.3\% in efficiency. 

The paper presents several significant contributions to the field of FL with a focus on robustness, fidelity, and efficiency. We summarize our contributions as follows.

\begin{itemize}
  \item \textbf{A novel Byzantine-robust aggregation method.} We propose \codename, containing two approaches, i.e., \agrmp and \agrxai. It achieves the triad of robustness, fidelity, and efficiency through the amplification of local updates.
  \item \textbf{A novel Aggregation Rule (\agr).} We design a distance-based \agr, which incorporates a top-up component of density measurement to supplement existing Euclidean distance and cosine similarity mechanisms. 
  \item \textbf{The seamless integration of \codename with existing \agrs.} We propose ten versions of 
  \codename that integrated three kinds of \agrs. 
  \item \textbf{A systematic evaluation.} We conduct a systematic evaluation on seven benchmark datasets against five poisoning attacks, demonstrating the notable enhancement of \codename over original Byzantine-robust methods across all experimented datasets.
\end{itemize}

%% file: chapters/background1.tex
\section{Background and Related Work}
\subsection{Federated Learning}
FL enables multiple clients to train a model collaboratively, which is worked through a central server that iteratively aggregates the local gradient updates computed by the clients.  
Specifically, let $r \in \left \{ 1, 2, \cdots, R  \right \}  $ be the current iteration of the training process, and $c_i$ be one client from $\mathcal{C}$ where $|\mathcal{C}|=N$. 
At $r=0$, the server initializes the global model and selects $\mathcal{C}_r \subseteq \mathcal{C}$ clients to broadcast the global model $\mathcal{W}_r$ to each client.
Then each client $c_i$  calculates the local updates $g_i^{[r]}$ based on its local datasets and uploads $g_i^{[r]}$ to the server. 
The server then applies an aggregation rule on the local updates to get the global model $\mathcal{W}_{r+1}$ for the next round of training. 

\subsection{Poisoning Attacks Against FL}
\IEEEpubidadjcol

Poisoning attacks can be divided into untargeted attacks \cite{cao_fltrust_2022,fang_local_2021,kurakin_adversarial_2017,li_lomar_2022,munoz-gonzalez_towards_2017,noauthor_manipulating_nodate} and targeted attacks \cite{bhagoji_analyzing_2019,bagdasaryan_how_2019} depending on the attacker's objectives. 
\begin{itemize}
    \item \textbf{Untargeted attacks.} Untargeted attacks aim to corrupt the global model to make incorrect predictions for any testing examples, therefore leading to undermining the test accuracy and integrity of FL models.
    One such technique involves manipulating training data and leading to a corrupted model (data poisoning attacks) \cite{biggio_poisoning_2012}. Another approach involves the manipulation of local updates directly (model poisoning attack) \cite{jagielski_manipulating_2018}. Recent research has also explored optimized and adaptive poisoning attacks, which seek to maximally perturb the reference aggregate in a manner that is detrimental to the model's performance while simultaneously evading detection by a Byzantine-robust aggregator \cite{cao_fltrust_2022}.
    \item \textbf{Targeted attacks.} Targeted attacks refer to a type of adversarial attack that is designed to target a subset of the training data and aims to influence the global model's predictions towards a particular class while maintaining overall prediction accuracy \cite{kairouz_advances_2021}. 
    For example, an attacker might inject specific patterns (trigger or backdoor) into the training data and change the label of those images with the trigger to a target class. In the inference stage, all images with the trigger will be mislabeled as the target class, while others will remain unaffected \cite{bagdasaryan_how_2019}.
\end{itemize}

\subsection{Byzantine-robust Aggregation Rules}
Dimension-wise Average \cite{dean_large_2012} is a useful \emph{aggregation algorithm (\agr)} to aggregate clients' gradients in non-adversarial FL settings. However, the Average \agr-based FL can be manipulated by malicious clients. Therefore, multiple Byzantine-robust \agrs \cite{cao_fltrust_2022,fang_local_2021,blanchard_machine_2017,mhamdi_hidden_2018,singhal_modern_nodate,eza_encyclopedia_2009,shen_better_2022,zhang2022fldetector,geng2023better,xu2022communication} are proposed to defend against poisoning attacks by malicious clients.

\begin{itemize}
\item \textbf{Distance-based mechanism.} 
Most distance-based \agrs rely on measuring the pairwise distance between local updates to identify and discard malicious updates. Krum and Multi-Krum~\cite{blanchard_machine_2017} identify one or $m$ local updates that are similar to others as the global model.
Trimmed Mean and Median \cite{yin_byzantine-robust_2018} employ approach involves coordinate-wise aggregation, where the \agr separately identifies and removes the outliers in each dimension and aggregates the remaining benign update.
Another approach, \emph{Adaptive federated average (AFA)} \cite{munoz2019byzantine}, compares the cosine similarity between the weighted average of collected gradients and each individual gradient. Gradients with cosine similarities outside a specified range are discarded, helping eliminate malicious gradients.

\item \textbf{Prediction-based mechanism.}
The prediction-based \agr test the gathered updates on a validation set to assess its prediction performance. The updates that lead to a decline in performance are subsequently removed before aggregation. One example is LoMar~\cite{li_lomar_2022}, which evaluates the quality of clients' model updates by analyzing the relative distribution of neighboring updates and determining an optimal threshold for differentiating between malicious and clean updates. Another representative mechanism is Fang~\cite{fang_local_2021}, which calculates losses and errors on the validation set of the updated model to determine whether the updates originate from malicious or benign clients. Additionally, Zhang \cite{zhang2022fldetector} identifies malicious clients by checking the consistency of their model updates. To elaborate, the server predicts a client's model update in each iteration based on historical model updates and flags a client as malicious if the received model update is inconsistent with the predicted update across multiple iterations.

\item \textbf{Trust bootstrapping-based mechanism.}
\emph{FLTrust} \cite{cao_fltrust_2022} is a  typical trust-bootstrapping based \agr. Each client is given a trust score based on the distance between the local updates and the reference gradients produced from a clean, trustworthy dataset. The trust score then serves as the weight when averaging updates.
\end{itemize}
The performance of the current Byzantine-robust techniques in terms of robustness, fidelity, and efficiency varies depending on the context. For instance, distance-based techniques, such as Krum \cite{blanchard_machine_2017} and Bulyan \cite{mhamdi_hidden_2018}, might not maintain robustness and fidelity against a large number of malicious participants. Fang \cite{fang_local_2021} may not be applicable to large-scale FL because it requires the extra computation cost for validating the loss of each individual update using the global model parameters. The integrity of trust bootstrapping-based approaches, such as FLTrust \cite{cao_fltrust_2022}, may be suppressed if the clean dataset deviates from the initial distribution. 

\subsection{Explainable Artificial Intelligence}

Explainable AI (XAI)~\cite{linardatos_explainable_2020,petsiuk_rise_2018,zeiler_visualizing_2013,lundberg_unified_2017,ribeiro_why_2016,burns_interpreting_2020,selvaraju_grad-cam_2020,simonyan_deep_2014,li_beyond_2019,sundararajan_axiomatic_2017,kindermans_learning_2017} is a collection of techniques that aim to increase the reliability and transparency of AI systems. 

Gradient-Weighted Class Activation Map (Grad-CAM) \cite{selvaraju_grad-cam_2020} is a widely used XAI technique, which is proposed to identify the importance of the image region by projecting back the weights of the output layer onto the convolutional feature maps. Specifically, it first computes the result of the gradient of the class score ($y^c$) w.r.t. the feature map in the last convolution layer ($A^{k}$), as $\frac{\partial y^{c}}{\partial A^{k}} $. Thus, for a given class $c$, it calculates the weights $\alpha_{k}^c$ corresponding to class $c$ for feature map $k$ as: 
\begin{equation}
\label{eq:weight}
\alpha_{k}^{c}=\frac{1}{Z} \sum_{i} \sum_{j} \frac{\partial y^{c}}{\partial A_{i, j}^{k}},
\end{equation}
where $Z$ denotes the number of pixels in the feature map, and $i$ and $j$ respectively index the width and height of the $k$-th feature map $A^{k}$.

%% file: chapters/method.tex
\section{\codename}
\subsection{Problem Statement}

\subsubsection{Attack Model}
We evaluate the proposed \codename method against untargeted and targeted attacks. 
Specifically, we consider an attack model which is in favor of the poisoning adversary among prior research~\cite{bagdasaryan_how_2019,bhagoji_analyzing_2019,cao_fltrust_2022,jagielski_manipulating_2018}
which allows the adversary to have the full knowledge of gradients generated by all participants, including those from benign ones. 
This enables the optimized and adaptive poisoning attacks and demonstrates compelling attack performance in recent studies~\cite{cao_fltrust_2022,jagielski_manipulating_2018}.
The attacker can also decide whether to inject malicious updates for the current round.
We assume the attacker does not compromise the central aggregator. 

\subsubsection{Defense Model}
We consider the defense strategy to be implemented on the server side. The server can gather a clean, small validation dataset for FL training. Since we only need a small clean dataset, e.g., 100 training examples, manual gathering and labeling on the server side is feasible. 

The defense aims to achieve Byzantine robustness against malicious clients and maintain fidelity and efficiency. Specifically, the strategy should not compromise the global model's classification accuracy, and it should be as accurate in non-adversarial conditions as the global model learned by FedAvg. Also, the method should retain computation efficiency and scalability in large-scale FL training.  

\subsection{\codename Overview}


In \codename, the server first collects local updates from participants and then extracts the most activated features (i.e., using \agrmp or \agrxai) to amplify the differences between benign gradients and malicious gradients. Then, the amplified gradients are concatenated for the following check. The detailed steps of \agrmp and \agrxai are as follows.

\subsubsection{\agrmp}

\begin{figure}
\vspace{-\baselineskip}
    \centering
    \includegraphics[width=3.5in]{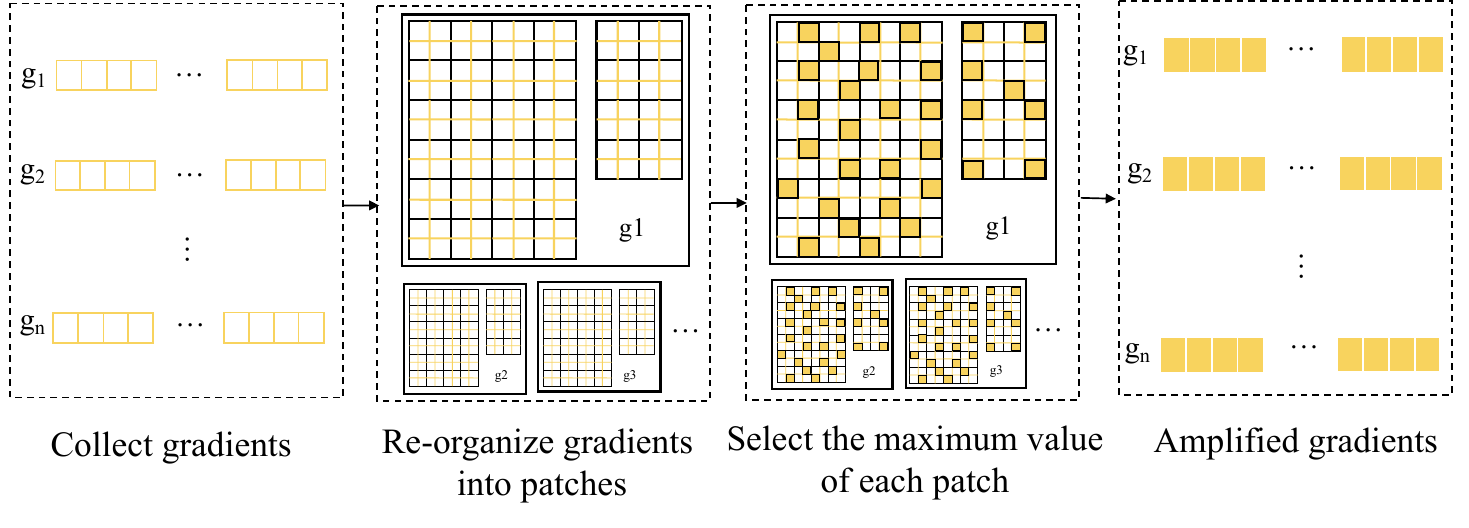}
    \caption{The amplification process on the collected gradients using \agrmp.}
    \label{fig:AGRMP}
\end{figure}

Fig. \ref{fig:AGRMP} and Algorithm \ref{alg:alg1} demonstrate the \agrmp method. Initially, the server acquires local updates $g_i$ from $N$ participants. Then, it divides each $g_i$ into patches using a kernel size of $k_p \times k_p$ to do the max filter. This involves computing the maximum value of each patch, which indicates the gradient of the most activated feature (function MAXFILTER in Algorithm \ref{alg:alg1}. After the max filter process, the amplified gradients are concatenated to facilitate cross-checking in the ensuing steps. 

\begin{algorithm}[h]
\renewcommand{\algorithmicrequire}{\textbf{Input:}}
\renewcommand{\algorithmicensure}{\textbf{Output:}}
\renewcommand{\algorithmicprocedure}{\textbf{function}}
\begin{algorithmic}[1]
\caption{\agrmp's amplification process.}
\label{alg:alg1}
\Require {
$g_i$ - received gradients from client $i$; 
$N$ - number of participants;
$k_p$ - the kernel size of each patch;
$H_{in}$ - the height of $g$;
$W_{in}$ - the width of $g$;
$Restore-size$ - whether restore the amplified gradients in the original size, default to being false.
}
\Ensure $G_{amp}$ the amplified gradients collection.
\Procedure{\agrmp}{$g_1,g_2,g_3,...$}
    \For{\texttt{$i = 1,2,3,...,N$}}
        
        \State {$g_{amp}^{i}\gets \textnormal{MAXFILTER}(g_i)$}        
            \If{$Restore-size$} 
                \State {Fill the dropped gradients in $g_{amp}^{i}$ with $0$s}  
            \EndIf 
    \EndFor
    \State {$G_{amp}\gets \left\{g_{a m p}^{i} \mid i=1,2, \ldots, N\right\}$} 
    \State \Return $G_{amp}$
\EndProcedure
\Procedure{MaxFilter}{$g$}

\State {$H_{out}\gets H_{in}/k_p$ } 
\State {$W_{out}\gets W_{in}/k_p$ }
    \For{\texttt{$h_o = 1,2,3,...,H_{out}$}}
        \For{\texttt{$w_o = 1,2,3,...,W_{out}$}}
        \State {$g_{amp} \leftarrow \max (\{g_{hi, wi} | hi \in [k_p*(ho-1)+1, k_p*ho], wi \in [k_p*(wo-1)+1, k_p*wo]$\}})
        \EndFor 
    \EndFor
    \State \Return $g_amp$
\EndProcedure
\end{algorithmic}
\end{algorithm}

\subsubsection{\agrxai}
\begin{figure}
    \centering
    \vspace{-1em}

    \includegraphics[width=3.5in]{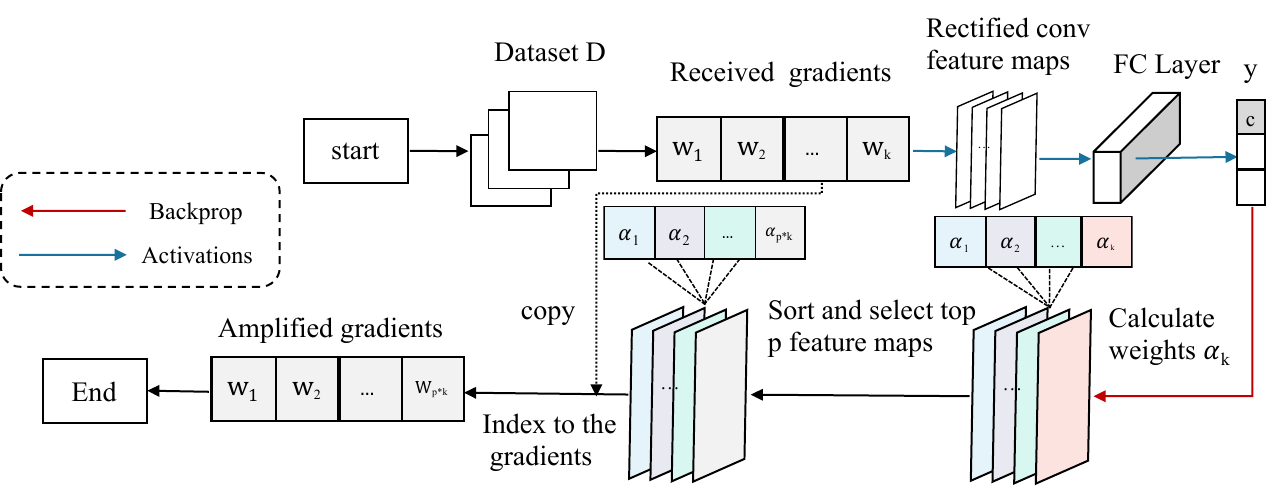}
    \caption{The amplification process on the collected gradients using \agrxai.}
    \label{fig:AGRXAI}
\end{figure}

Fig. \ref{fig:AGRXAI} and Algorithm \ref{alg:alg2} demonstrate the \agrxai method to extract the most activated gradients. Given that the last convolutional layer of the CNN typically has a satisfactory balance between high-level semantics and detailed spatial information, the gradient information flowing into this layer is utilized to understand the importance of each neuron for a specific decision of interest. 


Initially, a small clean training set $D$ is collected by the server (lines 4 in Algorithm \ref{alg:alg2}). After the server reviews gradients $g_i$ from $N$ clients, it will make a copy of the original gradients as $g_i^{og}$. The server then deploys $D$ to the received local model, where the gradient of the score w.r.t. the feature maps $A^k$ is computed (lines 6 in Algorithm \ref{alg:alg2}). These gradients flowing back are global-average-pooled over the width and height dimensions (indexed by i and j respectively) to obtain the importance weights $\alpha _k=\frac{1}{Z} \sum_{i} \sum_{j} \ \frac{\partial y}{\partial A_{i,j}^{k}}$ (lines 7 in Algorithm \ref{alg:alg2}). After getting the importance weights, we rank them in descending order and select the top $p$ feature maps with the highest weights. This selection allows us to focus on the most significant feature maps (need to note that \agrxai only focuses on the gradients of the last convolutional layer).
By indexing the gradients corresponding to these selected feature maps, represented as $G_{amp}$, we can utilize them for conducting the following defense check.

\begin{algorithm}
\renewcommand{\algorithmicrequire}{\textbf{Input:}}
\renewcommand{\algorithmicensure}{\textbf{Output:}}
\renewcommand{\algorithmicprocedure}{\textbf{function}}
\begin{algorithmic}[1]
\caption{\agrxai's amplification process.}\label{alg:alg2}
\Require {$g_i$ - received gradients from client $i$; 
$g_i^{og}$ - a copy of original gradients
$N$ - number of participants;
$D$ - server-side clean dataset; 
$\alpha_k$ - contribution weight of feature map $k$; 
$A^k$ - feature maps of a convolutional layer.}
\Ensure $G_{amp}$ the amplified gradients collection.
\Procedure{\agrxai}{$g_1,g_2,g_3,...$}
\For{\texttt{$i = 0,1,2,...,N$}}
    \State{$g_i^{og}$ $\gets$ copy $g_i$}
    \State Pass $D$ through the received model
    \State $y \gets$ final output logits 
    \State Backpropagate the weight $\frac{\partial y}{\partial A^{k}}$
      \State $\alpha _k\gets\frac{1}{Z} \sum_{i} \sum_{j} \ \frac{\partial y}{\partial A_{i,j}^{k}}$ 
      \State Sort and select the top $p$ feature maps
      \State $g_{amp}^{i}\gets$ index to the corresponding gradients stored in $g_i^{og}$
                \If{$Restore-size$} 
                    \State {Fill the dropped gradients in $g_{amp}^{i}$ with $0$s}  
                \EndIf 
        \State {$G_{amp}\gets \left\{g_{a m p}^{i} \mid i=1,2, \ldots, N\right\}$} 
\EndFor
    \State \Return $G_{amp}$
\EndProcedure




\end{algorithmic}

\end{algorithm}

\subsection{Equipping \codename with Byzantine-robust Aggregation Rules}

\begin{figure*}
\vspace{-\baselineskip}
    \centering
    \includegraphics[width=6in]{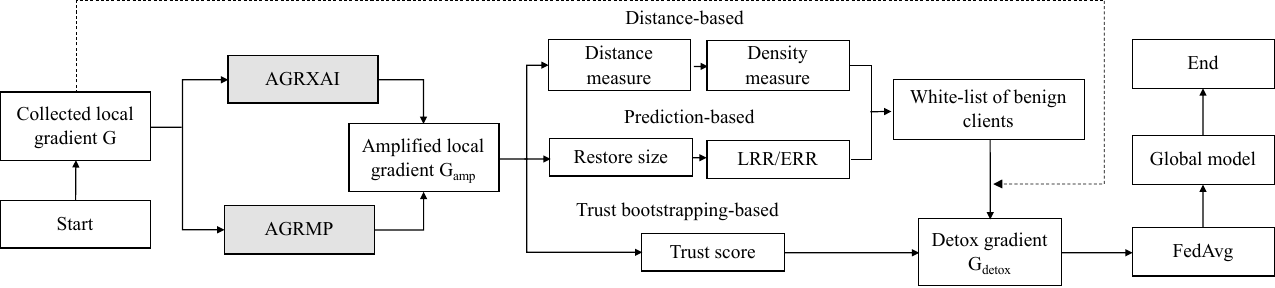}
    \caption{Workflow of \codename for Byzantine-robust mechanisms.}
    \label{fig:equip}
\end{figure*}

This section outlines the technical details of adapting \codename to support Byzantine-robust aggregators. The focus of the discussion is on three categories of aggregation mechanisms, and an overview of the workflow of \codename for Byzantine-robust mechanisms is provided in Fig. \ref{fig:equip}.
Specifically, the collected gradients $G$ are first amplified to $G_{amp}$ and used to calculate pairwise distances, loss value (LRR), and test accuracy (ERR), as well as trust scores for distance-based, prediction-based, and trust bootstrapping mechanisms, respectively. In the case of prediction-based mechanisms, the amplified gradients $G_{amp}$ are restored to their original size by filling the dropped gradients in $g_{amp}^{i}$ with $0$. The distance-based and prediction-based aggregators output a white list of benign participants denoted as $G_{detox}$, and the original gradients belonging to $G_{detox}$ are fed to FedAvg for global model computation. For trust bootstrapping-based mechanisms, it will first compute a trust score for each $g_i$, and the trust score then serves as the weight when averaging updates. Subsequent sections provide detailed information on \codename's implementation for each of the three categories of aggregation mechanisms.

\subsubsection{\textbf{\codename for Distance-based Aggregation Rules}}
Before aggregating the gradients, distance-based aggregation rules \cite{yin_byzantine-robust_2018,mhamdi_hidden_2018} examine the collected gradients and use distance measurements (e.g., Euclidean distance and cosine similarity) to measure the difference between malicious and benign ones. 
However, previous distance-based techniques primarily focus on anomaly detection, which becomes less effective as the number of malicious nodes rises. To address this issue, we design a \emph{density measurement component} to supplement the traditional distance-based mechanisms by using the knowledge that malicious gradients tend to be sparsely distributed, while benign ones are denser \cite{li_lomar_2022}. 
Our proposed method involves evaluating the density of the neighborhood surrounding each gradient. 
The neighborhood is defined as a set of $K$ neighbors, where $K$ must exceed half of the total number of participants $N$. The approach considers gradients residing in denser neighborhoods benign, whereas those in sparser areas are malicious. To implement this method, we first select the $K$ nearest neighbors of each participant's update based on distance measurement and record the scores of each measurement. Then, we sum up the score of the $K$ neighbors as the density score. The top $N_t$ participants with the highest density scores are then included in the whitelist. This density measurement is integrated into the distance-based mechanisms as a top-up component, illustrated in the right box in the upper part of Fig. \ref{fig:equip}, to improve the efficacy of the original approach.
\begin{figure}
\vspace{-\baselineskip}
\centering
    \subfloat[Untargeted attack]{\includegraphics[width=1.6in]{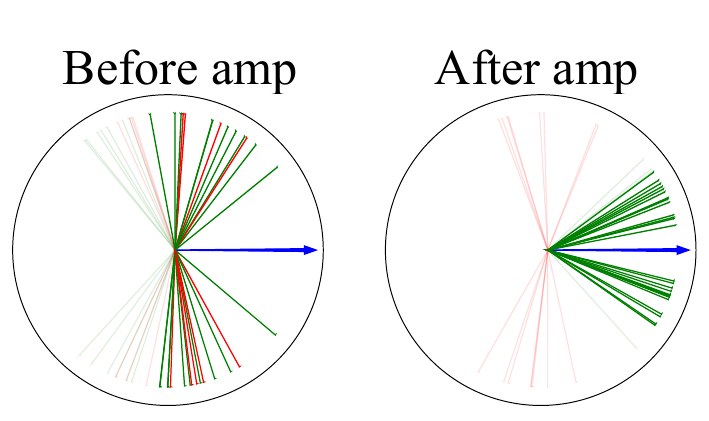}%
\label{fig:cos-before}}
\hfil
\hfil
\subfloat[Targeted attack]{\includegraphics[width=1.6in]{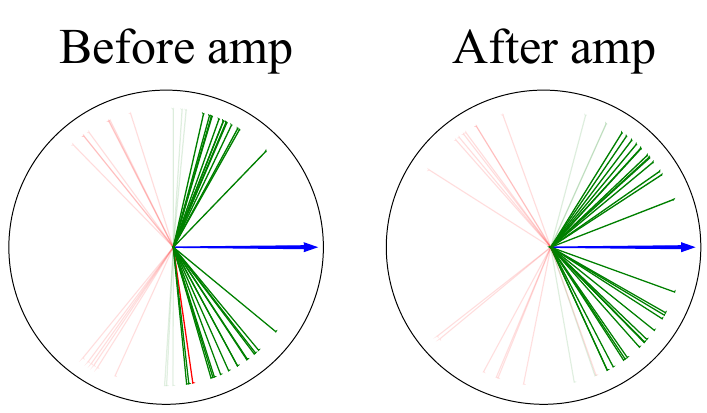}%
\label{fig:cos-after}}
\caption{A benign participant is selected (the blue arrow), and the $K$-nearest neighbors are determined using cosine similarity. The neighboring participants are further categorized as malicious (the red lines) or benign (the green lines). }
\label{fig:cos}
\end{figure}




Fig. \ref{fig:cos} demonstrates the effectiveness of the \codename mechanism, as observed in both targeted and untargeted attacks. The number of malicious participants in the neighborhood of a benign participant is notably reduced after amplification, as evidenced by a decrease from 8 to 0 for the untargeted attack and from 1 to 0 for the targeted attack. Additionally, the summed cosine similarity in the neighborhood undergoes a significant increase after amplification, with values escalating from 8.97 to 27.73 for the untargeted attack and from 13.86 to 21.73 for the targeted attack.
The pseudocode of \codename for Distance-based Aggregation Rules is displayed in Supplemental Document~A. 

\subsubsection{\textbf{\codename for Prediction-based Aggregation Rules}}
The Prediction-based mechanism \cite{fang_local_2021} applies the collected gradients to a validation dataset to evaluate the model's prediction performance in terms of LRR and ERR. It removes those clients degrading the model performance before aggrate the gradients.
The middle section of Fig. \ref{fig:equip} indicates how \codename works for a prediction-based mechanism. The amplification of the gathered parameters is first carried out by \codename. Then, we calculate the LRR and ERR, and those that perform better in prediction are added to the white list, which will be sent to global aggregation. 
The pseudocode of \codename for Prediction-based Aggregation Rules is displayed in Supplemental Document~B.

\subsubsection{\textbf{\codename for Trust Bootstrapping-based Aggregation Rules}}
This mechanism \cite{cao_fltrust_2022} utilizes a trustworthy clean dataset to generate clean gradients. By contrasting each participant's update with the clean gradients, a trust score is assigned to each participant. Then, the server utilizes the trust score as the weight to aggregate updates for the global model. 
The bottom section of Fig. \ref{fig:equip} indicates how \codename works for a Trust Bootstrapping-based mechanism. In addition to amplifying the collected gradients $g_i$, \codename also applies amplification to the gradient produced from the trust validation set. The trust score for each participant is calculated by utilizing the ReLU-clipped cosine similarity of the amplified gradients. The local model update data magnitudes are then normalized and combined to form the global model, weighted by their corresponding trust scores.
The pseudocode of \codename for Trust Bootstrapping-based Aggregation Rules is displayed in Supplemental Document~C. 

%% file: chapters/setup.tex
\section{Experimental Setup}
In this section, we conduct experiments to determine the triad of Byzantine-robust for the proposed method. 
\subsection{Datasets} 
We use seven real-world datasets for evaluation which are widely used in \agr literature. We follow previous work \cite{fang_local_2021} to distribute the training samples. 

\textbf{CIFAR-10} \cite{kindermans_learning_2017} consists of $60,000$ 
images, each sized by $32 \times  32$, divided into 10 classes. In our study, we employ pre-trained convolutional layers of ResNet56~\cite{he_deep_2016}, and the last three fully connected (FCN) layers have a size of \{64, 1024, 10\}. Each participant 
is provided with an IID version.

\textbf{MNIST} dataset \cite{shokri_privacy-preserving_2015} consists of $70,000$ grayscale images of handwritten digits, each with a size of $28 \times 28$ pixels. The dataset is divided into 10 classes representing the digits 0-9. For our experiments, we employ the FCN with layer sizes of \{784, 512, 10\}. Each participant in the study is provided with the an IID version.

\textbf{Fashion-MNIST} \cite{xiao_fashion-mnist_2017} is a 10-class fashion image for classification tasks, using a training set of $60,000$ images and a testing set of $10,000$ images with a resolution of 28x28 pixels. Here, we use the FCN with size \{784, 512, 10\} for our experiments, and each participant in the study is provided with an IID version.

\textbf{CATvsDOG Kaggle} \cite{noauthor_dogs_nodate} contains $25,000$ images of dogs and cats, each labeled accordingly.  We use the FCN with size \{2880, 512, 2\} for our experiments, and each participant in the study is provided with an IID version.

\textbf{PURCHASE100} \cite{shokri_membership_2017} contains $197,324$ records with 600 binary features of shopping histories 
and are classified into 100 distinct categories. For our experiments, we utilize the FCN with a size of \{600, 1024, 100\}. Notably, the dataset is observed to be non-IID, with the number of samples in each label varying between 106 and 5214. 

\textbf{LOCATION30} \cite{yang_participatory_2016} is comprised of mobile users' location ``check-ins" in the Foursquare social network \cite{shokri_membership_2017}. The dataset consists of $5,010$ data samples, each with 446 binary features, and is classified into 30 distinct categories. We utilize the FCN with a size of \{446, 512, 30\} for our experiments. Notably, the dataset is observed to be non-IID, with the number of samples in each label varying between 97 and 308. 

\textbf{TEXAS100} \cite{shokri_membership_2017} consists of $67,330$ records, each with $6,169$ binary features of hospital stay records, and are classified into 100 distinct categories \cite{shokri_membership_2017}. For our experiments, we employ the FCN with a size of \{6169, 1024, 100\}. Notably, the dataset is observed to be non-IID, as the number of samples in each label varied between 236 and 3046.

\subsection{Poisoning Attacks} 
We consider the \codename strategy against both untargeted and targeted attacks.
\begin{itemize}
\item
\textbf{Grad-Ascent attack \texttt{(\textbf\texttt{G-asc})}} is to revise the gradient towards the adversarial direction. Prior works  \cite{kurakin_adversarial_2017,munoz-gonzalez_towards_2017} have shown that manipulating the gradients can effectively degenerate the performance of the global model. 
\item
\textbf{Optimized and adaptive attack \texttt{(S\&H Attack)}} \cite{shejwalkar2021manipulating} is a cutting-edge Byzantine robust aggregation attack. The attack is formalized as an optimization problem to maximally perturb the reference aggregation in the malicious direction while being adaptive to avoid \agr detection. 
\item
\textbf{Label-flip attack \texttt{(L-flip)}} is a widely referenced poisoning attack \cite{cao_fltrust_2022}, where the adversary manipulates the local training data, intentionally switching the label for training data.
we utilize the setting as \cite{shen_better_2022}, which flips the label $y$ to $M-y-1$, where $M$ denotes the number of labels.

\item
\textbf{\texttt{<L-flip>+<\textbf\texttt{G-asc}>}} We propose a novel attack approach denoted as \textbf{\texttt{<L-flip>+<\textbf\texttt{G-asc}>}}, which combines the techniques of label flipping and gradient ascent attacks. This attack involves the incorporation of the malicious updates generated by both methods, resulting in a more potent and effective attack. Specifically, the \textbf{\texttt{<L-flip>+<\textbf\texttt{G-asc}>}} approach involves the summation of the perturbations generated by the label flipping and gradient manipulation techniques.
\item 
\textbf{Scale attack} \cite{cao_fltrust_2022} is a targeted attack that duplicates a certain percentage of targeted local training examples and alters their labels to the target label. Then, each malicious client calculates its local model update using the augmented local training data throughout each iteration of FL and sends it to the server. To amplify the impact of the malicious effect updates, the malicious clients scale up the gradient by a factor before updating them to the server.
Specifically, we use the setting as \cite{cao_fltrust_2022} and set scaling factor $\lambda =n$, where $n$ denotes the number of clients.
\item
\textbf{Distributed Backdoor Attack (\texttt{DBA})} \cite{xie2019dba} is a targeted attack in which the trigger pattern is equally split into $d$ parts and embedded into the local training data of $d$ malicious client groups. Specifically, we use the same setting as \cite{zhang2022fldetector} and split the trigger into four parts.
\end{itemize}
\begin{figure}
\centering
    \subfloat[Untargeted attack]{\includegraphics[width=1.4in]{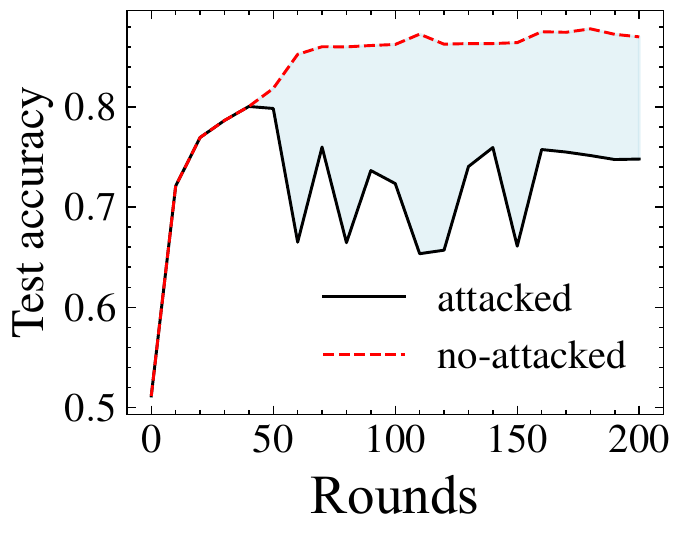}%
\label{fig:l}}
\
\subfloat[Targeted attack]{\includegraphics[width=1.4in]{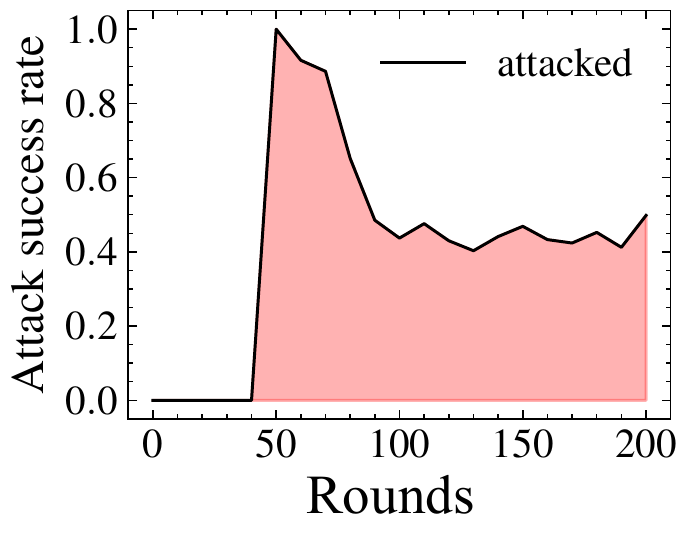}%
\label{fig:s}}
\caption{Illustration of the evaluation metrics.}
\label{fig:metrics}
\end{figure}

\subsection{Evaluated Defenses} \label{sec:evaldefense}
We introduce ten variations of \codename, which integrate three categories: distance-based mechanisms, prediction-based mechanisms, and trust bootstrapping-based mechanisms. For each category, we compare our \agrmp and \agrxai equipped versions it's base \agrs. 
\subsubsection{Distance-based Mechanisms}

\begin{itemize}
    \item \textbf{Cosine similarity, Euclidean distance, Merged distance with the Density measurement. }
    
    The evaluation focuses on density-based mechanisms, which are applied to the original gradients $G$ without any amplification. We propose two mechanisms: \cosden and \e, combining cosine similarity and Euclidean distance with Density measurement, respectively. We also introduce \m, which merges \cosden and \e by taking an intersection of their white lists. 

    \item \textbf {{\cosden}, {\e}, {\m} with \codename.}

    We evaluate {\cosden}, {\e}, {\m} equipped with \agrmp and \agrxai, which we refer to as {\cosmp}, {\emp}, {\mmp}, {\cosxai}, {\exai}, and {\mxai}. This allows us to assess the effectiveness of the mechanisms when applied to amplified gradients.
\end{itemize}

\subsubsection{Prediction-based Mechanisms}
\begin{itemize}
    \item \textbf{Fang's defence}. 
    We merge LRR and ERR for optimal performance in Fang's defense~\cite{fang_local_2021}, which serves as the based \agr for the prediction-based mechanism. 

    \item \textbf{Fang's defense with \codename.}
    We also evaluate Fang's defense equipped with \agrmp and \agrxai, which we refer to as {\fangmp} and {\fangxai}. In this case, the merged LRR and ERR are computed from the amplified gradients using \codename.
\end{itemize}

\subsubsection{Trust Bootstrapping-based Mechanisms}
\begin{itemize}
    \item \textbf{{\fl}}. We also evaluate {\fl}, which is a typical trust bootstrapping-based mechanism~\cite{cao_fltrust_2022}. To achieve better performance, 
    we set the trust set size as 2000 for CIFAR-10 \cite{kindermans_learning_2017}, MNIST \cite{shokri_membership_2017}, Fashion-MNIST \cite{xiao_fashion-mnist_2017}, CATvsDOG Kaggle \cite{noauthor_dogs_nodate}, PURCHASE100 \cite{shokri_membership_2017} and TEXAS100 \cite{shokri_membership_2017}, and 300 for the LOCATION30 dataset \cite{yang_participatory_2016}.

    \item \textbf{{\fl} with \codename}. Furthermore, we evaluate {\fl} equipped with \agrmp and \agrxai, which we refer to as {\flmp} and {\flxai}. In this case, the trust score is computed based on the amplified gradients using \codename.
\end{itemize}

\subsection{Evaluation Metrics}
We evaluate our \codename against targeted and untargeted attacks and take the distance-based, prediction-based, and trust bootstrapping-based aggregation rules as baselines to compare under the following evaluation metrics. 

For untargeted attacks, we use the \emph{test accuracy rate (TA)} of the global model to evaluate the performance. We denote $a_r$ as the TA at iteration $r$ without attack, while $\hat{a}_{r}$ denote the test accuracy at iteration $r$ under attack. The running time has a great impact on the TA \cite{shen_better_2022}. Therefore, we utilize the averaged TA over a monitoring period as the evaluation metric, which can also alleviate the impact of irrelevant factors such as overfitting on the performance evaluation. We calculate this averaged metric $\mathcal{L}$ as:
\begin{equation*} 
\mathcal{L}=\frac{1}{R_{1}-R_{0}} \sum_{r=R_{0}}^{R_{1}} a_{r}-\hat{a}_{r}.
\end{equation*}

We illustrate $\mathcal{L}$ as the blue shaded area (then divided by the round number) in Fig. \ref{fig:l}, which is generated on the CATvsDOG Kaggle dataset with the \textbf{\texttt{\textbf\texttt{G-asc}}} attack.

In targeted attacks, the goal is to maintain the test accuracy rate while inducing the global model to predict towards the target class. Therefore, we employ \emph{Attack Success Rate (ASR)} as the evaluation metric. 
The ASR is determined by analyzing the test data containing a trigger that is misclassified as the target class by the global model. 
By measuring the ASR, we are able to evaluate the extent to which the poisoning attack influences the global model's behavior and its ability to induce incorrect predictions toward the target class. 
We denote the ASR at iteration $r$ as Eq.~(\ref{eq:sr}):
\begin{equation}
    \label{eq:sr}
    s_{r}= \lVert \{ \hat{d} \mid_{\hat{d} \in \hat{D} , \omega_{r}(\hat{d})=c} \} \rVert / {\|\hat{D}\|},
\end{equation}
where $\hat{D}$ denote the malicious set, $\omega_{r}$ denote the global model at iteration $r$ and $c$ is the target label. Then we calculate the average ASR as Eq.~(\ref{eq:sa}) to evaluate the performance:
\begin{equation}
    \label{eq:sa}
S=\frac{1}{R_{1}-R_{0}} \sum_{r=R_{0}}^{R_{1}}\left(s_{r}\right).
\end{equation}
We illustrate $S$ as the red shaded area (then divided by the round number) in Fig. \ref{fig:s}, generated on the CATvsDOG Kaggle dataset with scale attack.
\subsection{FL Settings}

We train a five-layer CNN with two convolution layers, one max-pooling layer, and two dense layers. By default, each round involves the participation of $50$ clients. 
We employ the global communication rounds are $Rg = 200$. Each local client conducts $2$ epochs with a batch size of 64. The attacker initiates the attack at round $Rg = 50$, and we record test accuracy systematically every $10$ round. 

It's worth noting that in the default setting, we choose a higher malicious rate (i.e., 30\%) compared to previous works  \cite{fang_local_2021,cao_fltrust_2022} (around 20\%) to demonstrate the advantage of our method in defending against a stronger attack which compromises high proportion of malicious clients. In Section~\ref{sec:inf}, we also discuss the influence of the proportion of malicious clients, which ranges from $0.2$ to $0.4$. 

%% file: chapters/experiment.tex
\section{Experiment Results}
\label{sec:exp}
In this section, we discuss the performance of \codename: \cosmp, \emp, \mmp, \fangmp, \flmp, \cosxai, \exai, \mxai, \fangxai, \flxai against the five state-of-the-art poisoning attacks in terms of \emph{robustness}, \emph{fidelity} and \emph{efficiency} against targeted and untargeted attacks. The results show that \codename outperforms its base mechanisms on all three properties. 
 
In particular, \exai shows the highest robustness against targeted attacks, while \fangxai performs the best against untargeted attacks. Furthermore, the fidelity of \cosmp is superior, followed closely by \cosxai. 
Additionally, \flmp exhibits superior efficiency with small to medium network sizes, whereas \cosmp delivers the best efficiency with large networks. Now, we demonstrate a comprehensive performance analysis of \codename.

\subsection{Robustness Boosting}
\label{sec:robust}
\codename shows its capability to boost the robustness of untargeted and targeted attacks. We examine it as follows.
\subsubsection{Untargeted attack}
\begin{figure}
\vspace{-2em}
\centering
    \subfloat[Untargeted attack]{\includegraphics[width=1.3in]{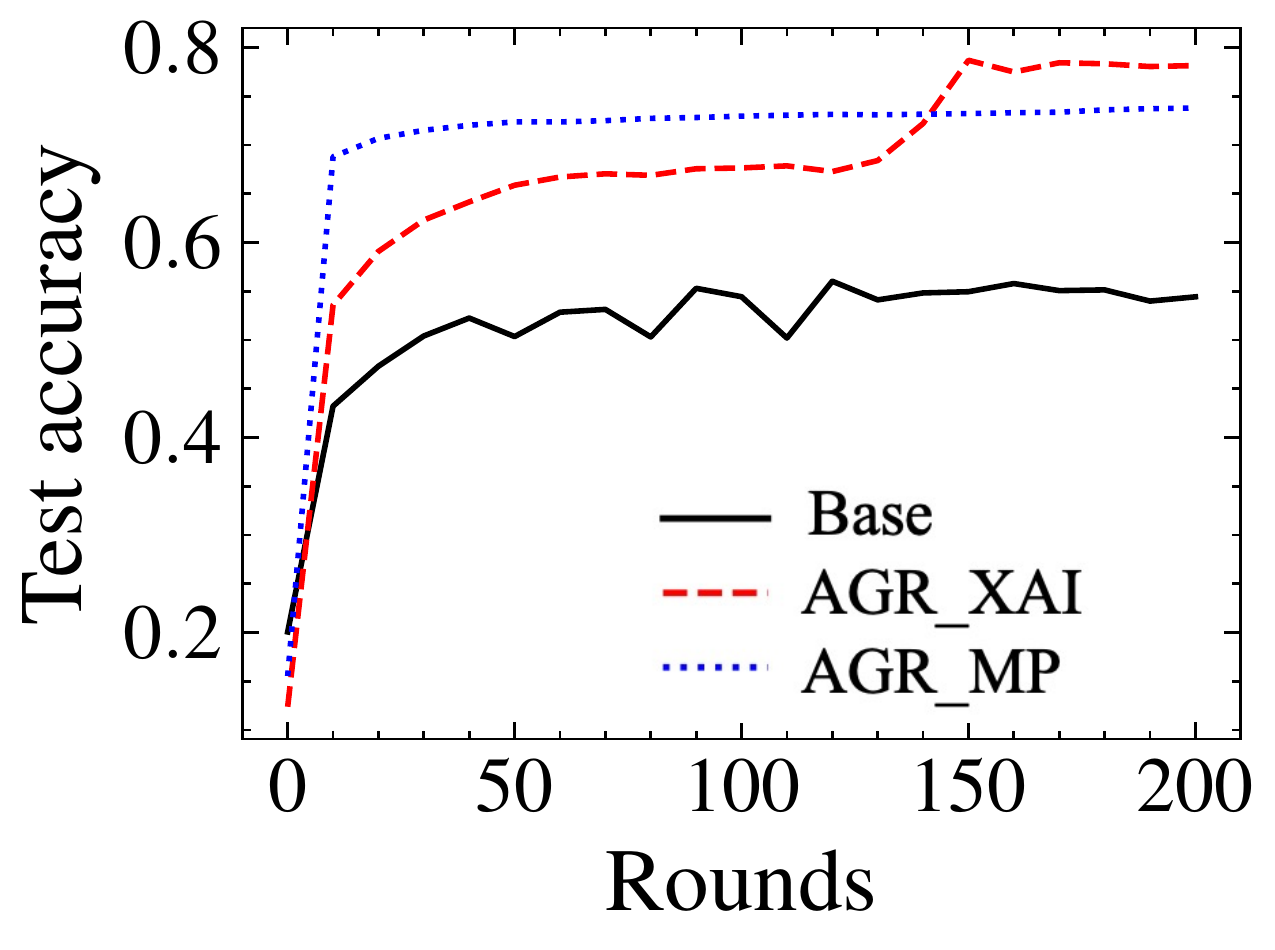}%
\label{fig:untargeted-xai}}
\hfil
\subfloat[Targeted attack]{\includegraphics[width=1.3in]{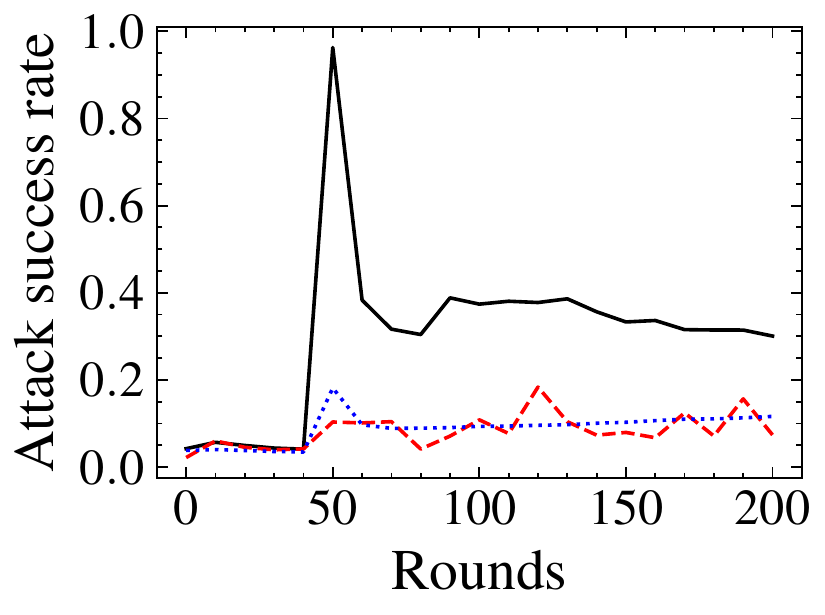}%
\label{fig:targeted-xai}}
\caption{Run-time TA for untargeted attacks and run-time ASR for targeted attacks.}
\label{fig:1}
\end{figure}

\begin{figure}[!t]
\vspace{-\baselineskip}
\centering

\subfloat[\cosden]{\includegraphics[width=1.1in]{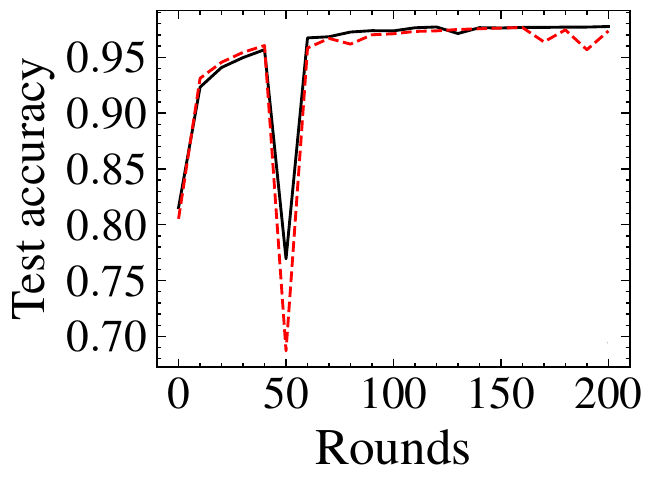}
\label{fig:6b}}
\hfil
\subfloat[\cosmp]{\includegraphics[width=1.1in]{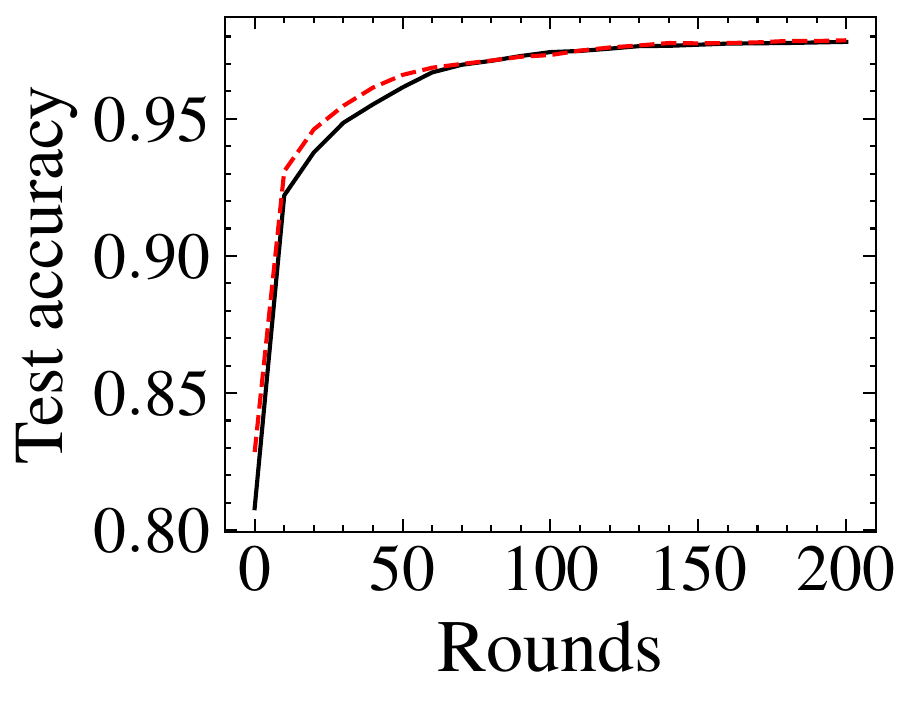}
\label{fig:6c}}
\hfil
\subfloat[\cosxai]{\includegraphics[width=1.1in]{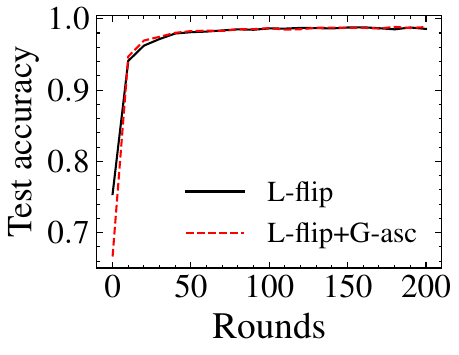}
\label{fig:6d}}
\caption{Run-time TA. (a): Negative pulse at
round 50 (the attack starts round). (b) (c): \codename mitigates the negative pulse. }
\label{fig:negative}
\end{figure}

The \codename demonstrates its ability to improve the robustness of the FL system in general, as evidenced by the results presented in Fig. \ref{fig:untargeted-xai}. Specifically, the \codename-equipped \agr achieves higher TA than its counterpart, outperforming it by 45.21\%. These findings are supported by the experimental results presented in Table \ref{table:overall}. Specifically, \agrmp outperforms its counterpart base aggregator in distance-based, prediction-based, and trust bootstrapping-based mechanisms with 62.21\%, 35.63\%, and 16.19\%, respectively. Similarly, \agrxai exhibits superior performance over its base aggregator in distance-based, prediction-based, and trust bootstrapping-based mechanisms by 88.68\%, 2.41\%, and 71.11\%, respectively.
We further report several observations.
\paragraph{Negative pulse mitigation}\label{sec:negative_pulse}
In FL training, the presence of these malicious actors can cause significant disruptions to the training process and can severely degrade the accuracy and performance of the shared model. One of the observed phenomena associated with such attacks is the ``negative pulse", which is characterized by an abrupt reduction in the model's TA caused by the poisoning adversary at the early stage of the attacks. This phenomenon is consistently observed across different datasets and FL frameworks, indicating the need for effective defense mechanisms to mitigate it.
We observe that \codename shows promise in effectively mitigating negative pulses as shown in Fig. \ref{fig:negative}.



\paragraph{Base mechanism matters}
Also, it should be noted that the efficacy of the \codename-equipped version is predicated on the effectiveness of its base aggregator. If the base aggregator exhibits a weakness in a specific context, this vulnerability may propagate to the amplified version. For instance, the distance-based \cosmp defense demonstrates an $\mathcal{L}$ error rate of 16.95\% against \textbf{\texttt{L-flip}} attacks on LOCATION30. While the \cosmp defense improves its performance, achieving an $\mathcal{L}$ error rate of 5.96\%, it still falls behind other defenses that achieve an average $\mathcal{L}$ error rate of 0.28\%. 
\subsubsection{Targeted attack}
Targeted attacks in FL show a significant challenge for detection due to the subtle and inconspicuous nature of the triggers that are designed to blend in with normal model updates. However, our \codename exhibits promising performance in detecting targeted attacks. Specifically, Fig. \ref{fig:targeted-xai} presents the results of a targeted attack against base aggregators and \codename, demonstrating that \agrxai outperforms its base aggregator counterpart by achieving a 34.94\% lower $S$. Specifically, \agrmp outperforms its counterpart base aggregator in distance-based, prediction-based, and trust bootstrapping-based mechanisms with 25.02\%, 48.4\%, and 5.5\%, respectively. Similarly, \agrxai exhibits superior performance over its base aggregator in distance-based, prediction-based, and trust bootstrapping-based mechanisms by 59.22\%, 18.89\%, and 39.70\%, respectively.

It has been observed that the \agrxai-based \codename outperforms the \agrmp-based \codename, indicating the potential of XAI techniques in improving targeted attack detection in FL. Specifically, this enhanced performance can be attributed to \agrxai's ability to capture the significant features of an image that has a high probability of containing the trigger. By selectively manipulating these features, \agrxai is able to effectively detect the presence of triggers in model updates. These findings highlight the potential of XAI techniques for improving targeted attack detection in FL and suggest that \agrxai may be a promising approach for enhancing the robustness of FL models against targeted attacks.

The \textbf{\texttt{Scale}} attack and \textbf{\texttt{DBA}} attack against each defense mechanism for seven benchmark datasets are presented in Table \ref{table:overall}. Among the evaluated defenses, \fangxai exhibits consistently strong defense performance across all datasets, followed by \fangmp, which demonstrates effective defense against non-image datasets. However, \flmp shows an abnormal performance on TEXAS100, with a 63.76\% ASR for \textbf{\texttt{Scale}} attacks. This can be attributed to the fact that the trust bootstrapping-based mechanism heavily relies on the representativeness of the validation set.
Section \ref{sec:validation} presents a detailed evaluation of how the distribution of the validation dataset can impact the performance of our proposed method.


\vspace{-\baselineskip}

\begin{table}[]
\fontsize{25}{26}\selectfont 
\hspace{-10pt}

\renewcommand{\arraystretch}{1.2}
\centering

\caption{Averaged TA loss ($\mathcal{L}$) for untargeted attacks and ASR ($S$) for targeted attacks (presented in percentage). The best defense for each attack (row-wise) is in bold.}
\label{table:overall}

\resizebox{0.5\textwidth}{!}{

\begin{tabular}{c|c|ccccccccc|ccc|ccc}
 \toprule
 
\multirow{2}{*}{Attack} & \multirow{2}{*}{No def} & \multicolumn{9}{c}{Distance-based}                                         & \multicolumn{3}{|c}{Prediction-based} & \multicolumn{3}{|c}{Trust-based} \\
                        &                         & C     & E     & M    & CM    & EM    & MM    & CX    & EX    & MX    & F        & FM       & FX       & T      & TM      & TX     \\
\midrule
\midrule

\multicolumn{17}{l}{\textbf {Untargeted attacks}}                                                                                                                                             \\
\midrule
\multicolumn{17}{l}{CIFAR-10}                                                                                                                                                         \\
\textbf{\texttt{G-asc}}                   & 1.12                    & 0.71  & 1.70  & 0.23 & \textbf{-0.03} & -0.02 & 0.07  & 1.07  & 0.21  & 1.06  & 0.07     & 0.38     & 1.02     & 1.44   & 1.56    & 1.92   \\
\textbf{\texttt{S\&H}}                    & 0.12                    & 0.07  & -0.09 & 0.23 & -0.10 & 0.12  & \textbf{-0.18} & 0.01  & 0.09  & 0.23  & -0.11    & 0.11     & 0.12     & 1.96   & 1.45    & 1.01   \\
\textbf{\texttt{L+G}}                     & 2.87                    & 5.88  & 2.95  & 1.70 & 0.36  & 4.02  & 0.41  & 1.70  & \textbf{-1.52} & 1.25  & 0.50     & 0.51     & 1.74     & 2.03   & 1.82    & 1.18   \\
\midrule
\multicolumn{17}{l}{MNIST}                                                                                                                                                            \\
\textbf{\texttt{G-asc}}                   & 1.02                    & 1.33  & 0.24  & 0.27 & 0.23  & 0.01  & 0.14  & 0.07  & \textbf{-0.06} & 0.30  & 0.21     & 0.21     & 1.72     & 5.63   & 2.27    & 0.62   \\
\textbf{\texttt{S\&H}}                    & 0.41                    & 0.40  & 0.41  & 0.30 & 0.11  & 0.29  & 0.36  & 0.88  & 0.82  & \textbf{-0.17} & 0.14     & 0.14     & 0.03     & 1.70   & 2.18    & 1.27   \\
\textbf{\texttt{L+G}}                     & 0.73                    & 2.63  & 85.86 & 2.23 & 1.39  & 0.38  & 0.16  & 0.14  & \textbf{-1.06} & -0.58 & -0.08    & -0.08    & 0.10     & 5.60   & 2.35    & -0.54  \\
\midrule
\multicolumn{17}{l}{CAT-DOG}                                                                                                                                                          \\
\textbf{\texttt{G-asc}}                   & 6.80                    & 6.92  & 4.36  & 4.02 & 2.65  & 5.36  & 1.94  & 8.91  & 3.04  & 3.09  & 6.14     & 9.67     & 6.42     & 5.60   & \textbf{-0.48}   & 0.93   \\
\textbf{\texttt{S\&H}}                    & 3.19                    & 4.28  & 4.65  & 3.92 & \textbf{-2.83} & 2.06  & -0.21 & -0.18 & 0.50  & -2.58 & 1.74     & -0.55    & -0.54    & 8.43   & 5.54    & -0.13  \\
\textbf{\texttt{L+G}}                     & 28.54                   & 0.07  & 3.42  & 5.45 & 2.89  & -2.04 & 1.49  & -3.22 & \textbf{-4.69} & 3.46  & 1.32     & 0.14     & 1.32     & 7.18   & -3.32   & 0.08   \\
\midrule
\multicolumn{17}{l}{FASHION-MNIST}                                                                                                                                                    \\
\textbf{\texttt{G-asc}}                   & 5.02                    & 5.02  & 3.20  & 3.02 & 2.55  & 4.29  & 1.90  & 1.07  & 0.19  & 0.90  & 4.42     & 9.77     & 3.92     & 2.89   & \textbf{-0.12}   & 1.22   \\
\textbf{\texttt{S\&H}}                    & 2.10                    & 3.09  & 3.29  & 2.08 & 2.01  & 2.09  & -0.31 & 2.88  & \textbf{-3.82} & 0.27  & 1.59     & 0.01     & 0.83     & 5.09   & -0.42   & 2.27   \\
\textbf{\texttt{L+G}}                     & 20.04                   & 2.61  & 4.98  & 2.01 & 2.01  & 0.32  & 0.21  & \textbf{-3.14} & 1.02  & -0.28 & 2.09     & 0.12     & 0.60     & 6.47   & 2.31    & 0.34   \\
\midrule
\multicolumn{17}{l}{LOCATION30}                                                                                                                                                       \\
\textbf{\texttt{G-asc}}                   & 8.48                    & 16.95 & \textbf{0.50}  & 3.77 & 5.98  & 1.89  & 3.14  & -     & -     & -     & 3.40     & 4.28     & -        & 5.74   & 8.67    & -      \\
\textbf{\texttt{S\&H}}                    & 0.69                    & -0.24 & 1.31  & 2.76 & \textbf{-0.65} & 0.02  & 0.20  & -     & -     & -     & 2.40     & 1.35     & -        & 4.26   & 0.50    & -      \\
\textbf{\texttt{L+G}}                     & 6.50                    & 3.17  & 11.34 & 9.37 & \textbf{-0.45} & 6.44  & 3.23  & -     & -     & -     & 1.33     & 1.86     & -        & 2.74   & 8.36    & -      \\
\midrule
\multicolumn{17}{l}{PURCHASE100}                                                                                                                                                      \\
\textbf{\texttt{G-asc}}                   & 12.82                   & 17.49 & 17.61 & 1.64 & 0.96  & 2.25  & 2.52  & -     & -     & -     & \textbf{0.75}     & 1.04     & -        & 2.69   & 2.55    & -      \\
\textbf{\texttt{S\&H}}                    & \textbf{0.04}                    & 0.34  & 1.34  & 1.81 & 0.16  & 1.54  & 1.69  & -     & -     & -     & 0.66     & 0.89     & -        & 3.52   & 1,87    & -      \\
\textbf{\texttt{L+G}}                     & 33.78                   & 14.22 & 19.22 & 2.75 & \textbf{0.54}  & 1.07  & 0.75  & -     & -     & -     & 1.06     & 1.48     & -        & 4.82   & 2.64    & -      \\
\midrule
\multicolumn{17}{l}{TEXAS100}                                                                                                                                                         \\
\textbf{\texttt{G-asc}}                   & 1.98                    & 1.39  & 1.17  & 2.64 & 1.17  & 1.58  & 1.49  & -     & -     & -     & 2.00     & 0.52     & -        & 1.66   & \textbf{-0.15}   & -      \\
\textbf{\texttt{S\&H}}                    & 1.46                    & 0.58  & 1.87  & 2.51 & 0.53  & 1.91  & 1.99  & -     & -     & -     & 1.93     & 0.51     & -        & 0.51   & \textbf{-0.45}   & -      \\
\textbf{\texttt{L+G}}                     & 2.33                    & 3.23  & 5.52  & 5.20 & \textbf{1.25}  & 4.00  & 2.79  & -     & -     & -     & 2.40     & 1.96     & -        & 2.20   & 1.39    & -     \\
\midrule
\multicolumn{17}{l}{\textbf{Targeted attacks}}                                                                                                                                                         \\
\midrule
\multicolumn{17}{l}{CIFAR-10}                                                                                                                                                          \\
\textbf{\texttt{Scale}}                   & 94.44                   & 17.35 & 14.60 & 10.03 & 2.64  & 1.89  & 1.09  & 2.81  & \textbf{0.25}  & 1.74  & 49.60    & 2.70     & 0.49     & 12.25   & 9.91   & 4.21   \\
DBA                   & 85.64                   & 12.64 & 10.43 & 15.01 & 9.03  & 9.22  & 8.61  & 8.99  & 6.00  & 3.95  & 15.12    & 5.32     & \textbf{3.88}     & 13.29   & 18.00  & 13.38  \\
\midrule
\multicolumn{17}{l}{MNIST}                                                                                                                                                             \\
\textbf{\texttt{Scale}}                   & 16.58                   & 0.37  & 0.89  & 2.19  & 0.57  & 0.37  & 0.39  & 2.81  & 0.25  & 1.74  & 5.54     & \textbf{0.24}     & 0.49     & 3.59    & 0.35   & 0.79   \\
\textbf{\texttt{DBA}}                   & 16.95                   & 2.95  & 2.41  & 1.96  & 0.34  & 10.09 & 0.63  & 7.83  & 0.90  & 0.42  & 5.39     & 1.27     & \textbf{0.23}     & 3.20    & 0.46   & 1.80   \\
\midrule
\multicolumn{17}{l}{CAT-DOG}                                                                                                                                                           \\
\textbf{\texttt{Scale}}                   & 80.26                   & 37.14 & 27.90 & 26.13 & 27.14 & 16.40 & 16.13 & 16.18 & 14.36 & 12.50 & 24.23    & 14.03    & 10.68    & 20.53   & 10.30  & \textbf{6.88}   \\
\textbf{\texttt{DBA}}                   & 95.23                   & 24.47 & 23.53 & 22.24 & 14.03 & 14.37 & 13.83 & 12.24 & 13.28 & 17.63 & 19.52    & 10.43    & 20.07    & 12.79   & 10.39  & \textbf{10.23}  \\
\midrule
\multicolumn{17}{l}{FASHION-MNIST}                                                                                                                                                     \\
\textbf{\texttt{Scale}}                   & 79.95                   & 22.76 & 19.85 & 14.29 & 2.70  & 9.85  & 4.12  & 14.70 & 13.96 & 3.44  & 12.45    & \textbf{2.41}     & 14.3     & 14.39   & 4.83   & 3.20   \\
\textbf{\texttt{DBA}}                   & 80.01                   & 13.32 & 14.34 & 17.95 & 3.89  & 4.51  & 4.18  & 15.02 & 12.09 & \textbf{0.22}  & 23.21    & 3.32     & 10.32    & 13.45   & 3.24   & 2.33   \\
\midrule
\multicolumn{17}{l}{LOCATION30}                                                                                                                                                        \\
\textbf{\texttt{Scale}}                   & 53.45                   & 3.13  & 7.32  & 2.34  & 8.46  & 12.6  & 13.6  & -     & -     & -     & 5.54     & \textbf{1.39}     & -        & 3.59    & 4.68   & -      \\
\textbf{\texttt{DBA}}                   & 65.32                   & 3.21  & 3.66  & 2.40  & 2.74  & 2.98  & \textbf{1.93}  & -     & -     & -     & 7.43     & 2.34     & -        & 6.23    & 3.56   & -      \\
\midrule
\multicolumn{17}{l}{PURCHASE100}                                                                                                                                                       \\
\textbf{\texttt{Scale}}                   & 74.78                   & 0.56  & 3.21  & 2.67  & 1.73  & 2.82  & \textbf{0.04}  & -     & -     & -     & 0.05     & 1.02     & -        & 0.41    & 1.11   & -      \\
\textbf{\texttt{DBA}}                   & 69.03                   & 4.21  & 3.23  & 1.35  & 3.89  & 4.51  & 4.18  & -     & -     & -     & 2.23     & 3.12     & -        & 4.26    & \textbf{1.23}   & -      \\
\midrule
\multicolumn{17}{l}{TEXAS100}                                                                                                                                                          \\
\textbf{\texttt{Scale}}                   & 99.03                   & 11.16 & 32.02 & 12.43 & 33.1  & 14.4  & 0.03  & -     & -     & -     & \textbf{0.01}     & 0.03     & -        & 35.59   & 63.76  & -      \\
\textbf{\texttt{DBA}}                   & 89.21                   & 13.35 & 12.32 & 10.78 & 21.02 & 12.65 & 3.54  & -     & -     & -     & \textbf{2.01}     & 3.12     & -        & 34.04   & 15.54  & -       \\ 
\bottomrule                              
\end{tabular}
}
\begin{flushleft}
    \scriptsize
    \textsuperscript{$\dagger$} In this table, we use No def to stand for no defense, 
    {C} to stand for \cosden, {CM} for \cosmp, {CX} for \cosxai,
    {E} to stand for \e, {EM} for \emp, {EX} for \exai,
    {M} to stand for \m, {MM} for \mmp, {MX} for \mxai,
    {F} to stand for \fang, {FM} for \fangmp, {FX} for \fangxai,
    {T} to stand for \fl, {TM} for \flmp, {TX} for \flxai
    (The abbreviation also applies to Tables \ref{table:fidelity}, \ref{table:TIME}, and \ref{table:bais}).
    \end{flushleft}
\end{table}

\subsection{Fidelity Boosting}
\label{sec:Fidelity}
\begin{table}[h]
\fontsize{25}{26}\selectfont
\hspace{-10pt}
\renewcommand{\arraystretch}{1.2}
\centering
\caption{Performance loss $\mathcal{L}$ in terms of fidelity.}
\label{table:fidelity}
\resizebox{0.5\textwidth}{!}{

\begin{tabular}{c|ccccccccc|ccc|ccc}
\toprule
\multirow{2}{*}{Datasets} & \multicolumn{9}{c|}{Distance-based}                                   & \multicolumn{3}{c|}{Prediction-based} & \multicolumn{3}{c}{Trust-based} \\
                         & C     & E    & M    & CM    & EM   & MM   & CX   & EX   & MX   & F        & FM       & FX       & T      & TM      & TX     \\
\midrule
CIFAR-10                 & -0.07 & 0.50 & 0.68 & 0.08  & 0.01 & 0.24 & 0.17 & 0.27 & 0.29 & 0.15     & 0.23     & 0.29     & 1.48   & 1.38    & 1.29   \\
MNIST                    & 0.41  & 0.74 & 0.62 & 048   & 0.72 & 0.88 & 0.14 & 0.13 & 0.10 & 0.64     & 0.54     & 0.15     & 0.62   & 1.39    & 0.13   \\
CAT-DOG                  & 0.43  & 0.42 & 0.44 & 0.42  & 0.41 & 0.40 & 0.42 & 0.45 & 0.41 & 0.38     & 0.39     & 0.38     & 0.48   & 0.62    & 0.42   \\
FASHION-MNIST            & 0.51  & 0.63 & 0.55 & 0.46  & 0.39 & 0.23 & 0.21 & 0.30 & 0.24 & 0.34     & 0.25     & 0.21     & 0.35   & 0.43    & 0.20   \\
LOCATION30               & 0.00  & 2.84 & 3.59 & -0.15 & 1.35 & 1.69 & -    & -    & -    & 3.64     & 1.29     & -        & 3.21   & 1.87    & -      \\
PURCHASE100              & 0.53  & 1.23 & 1.69 & 0.28  & 1.48 & 1.29 & -    & -    & -    & 0.69     & 0.94     & -        & 3.17   & 1.98    & -      \\
TEXAS100                 & 0.68  & 0.98 & 1.49 & 0.18  & 1.48 & 1.18 & -    & -    & -    & 1.30     & 0.38     & -        & 0.31   & -0.53   & -   \\
\bottomrule
\end{tabular}

}
\end{table}

\begin{figure}
\vspace{-\baselineskip}
    \centering
    \includegraphics[height=1.5in]{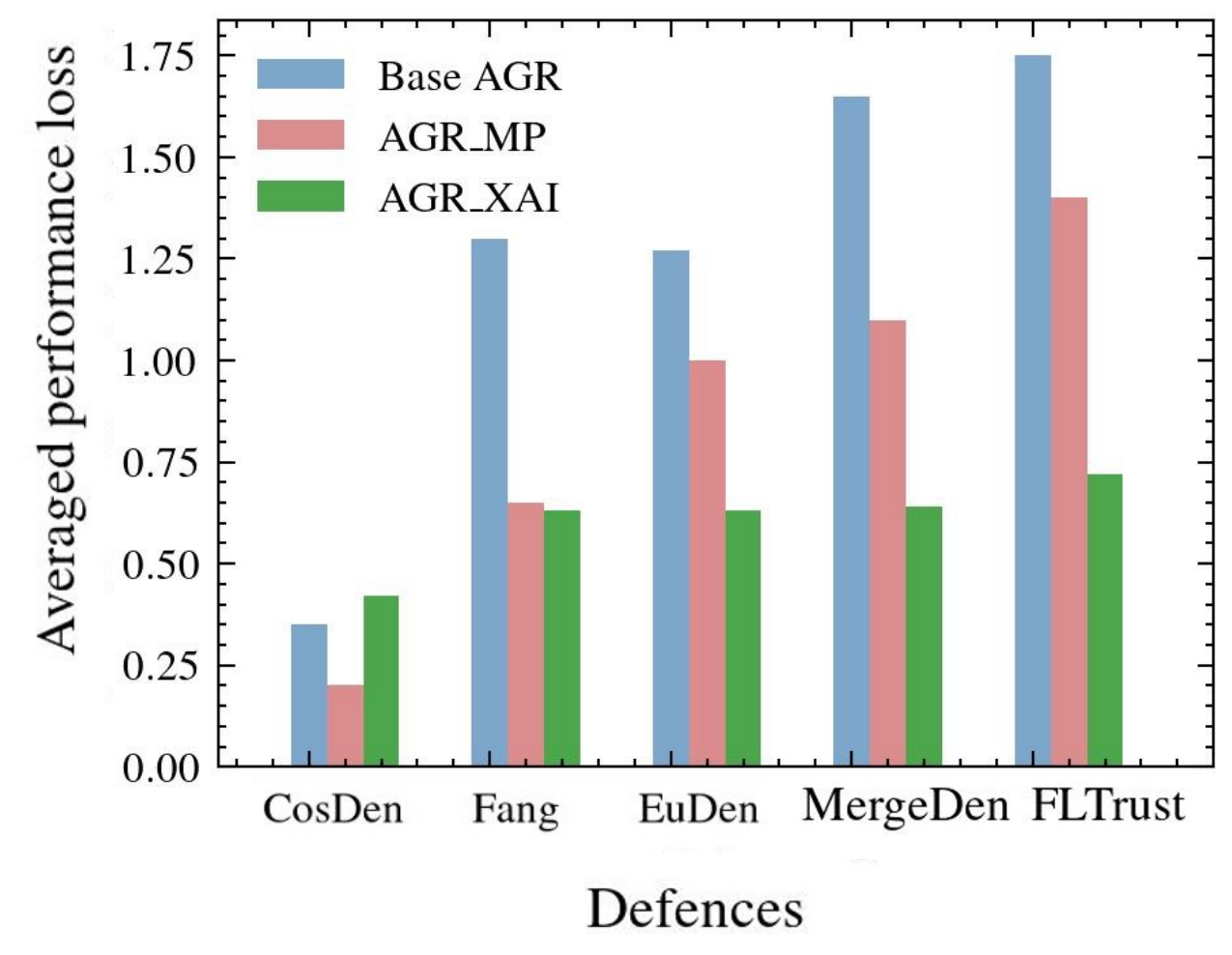}
    \caption{Averaged performance loss $\mathcal{L}$ across all datasets in terms of fidelity.}
    \label{fig:fidelity}
\end{figure}

Fidelity is a crucial factor in assessing an aggregator's effectiveness in preserving helpful information during the FL process. 
Fig. \ref{fig:fidelity} presents a comparison of the fidelity scores between \codename and their base \agrs. Our results demonstrate that \codename (indicated by the pink and green bars) outperforms the original base aggregators (indicated by the blue bars) by showing significantly less fidelity loss. These findings suggest that \codename is effective in improving the fidelity of base aggregators, which is essential for preserving the quality of the trained model and enhancing the robustness of the federated learning process.

Our evaluation reveals that the distance-based \cosmp aggregator performs the most promisingly, achieving an average fidelity loss of 0.18\%. Furthermore, the prediction-based \fangxai aggregator also demonstrates desirable fidelity with an average loss of 0.68\%. However, the performance of trust bootstrapping-based \flxai highly depends on the dataset used. Notably, this approach can significantly reduce the fidelity loss in some datasets. Still, it may yield a higher fidelity loss on others, such as LOCATION30 and PURCHASE100, where the validation dataset distribution differs from the original training set, potentially leading to biased global models. The detailed results
are given in Table \ref{table:fidelity}.
\subsection{Efficiency Boosting}
\label{sec:Efficiency}
\begin{table}[]
\fontsize{20}{21}\selectfont
\hspace{-10pt}
\renewcommand{\arraystretch}{1.2}
\centering

\caption{Time consumption (in second) of aggregation on CIFAR-10.}
\label{table:TIME}
\resizebox{0.5\textwidth}{!}{
\begin{tabular}{c|ccc|ccc|ccc}
 \toprule
Hidden layer & C     & CM    & CX     & F     & FM    & FX     & T     & TM    & TX     \\
\midrule
512          & 330.8 & 241.0 & 2128.4 & 439.8 & 341.1 & 2383.3 & 316.4 & 233.2 & 2340.8 \\
1024         & 328.6 & 240.6 & 2308.1 & 426.9 & 340.2 & 2352.7 & 324.8 & 233.1 & 2401.1 \\
2048         & 374.3 & 306.8 & 3542.3 & 442.6 & 412.3 & 2342.8 & 320.2 & 295.3 & 2538.7 \\
\bottomrule
\end{tabular}

}

\end{table}

Table \ref{table:TIME} presents the time consumption of \codename experiments carried out on CIFAR-10 utilizing the NVIDIA GeForce RTX 3080 device. 
For \agrmp-based methods, 
the time complexity of the amplification process is $\mathcal{O}(N*P_n)$  in each iteration, (where $N$ is the number of clients and $P_n$ is the number of parameters in the model). 
Since the value of $P_n$  is significantly reduced by the amplification, our \agrmp effectively lowers the computational cost for \agr, particularly for large-size neural networks. 
Specifically, as is shown in Table \ref{table:TIME}, \agrmp reduces total time consumption by 20\% from 369.38 seconds to 293.73 seconds, on average.
However, \agrxai-based methods exhibit increased time consumption due to the additional computation involved in collecting feature weights. The time complexity of the \agrxai amplification process is $\mathcal{O}(N*(N_c+P_n^2))$, where $N_c$ is the total number of computations involved in the forward pass and gradient calculation \cite{selvaraju_grad-cam_2020}.

The space complexity of \codename is equivalent to that of the traditional FL method, i.e., $\mathcal{O}(N*P_n)$, which exhibits linear growth with respect to both $N$ and $P_n$. \\

\subsection{Influence Factors}
\label{sec:inf}
This section presents an evaluation of \codename's performance in varying scenarios concerning 1) the number of participants, 2) the malicious factor, and 3) the proportion of the extracted gradient. Specifically, in Section \ref{sec:partiNumber}, we consider 30, 50, and 100 participants while fixing the percentage of malicious participants at 30\%. In Section \ref{sec:maliPortion}, we consider 20\%, 30\%, and 40\% proportions of malicious participants, with a fixed number of participants at 50.  In Section \ref{sec:top}, we consider the kernel size to be within the range of 2×2, 3×3, 5×5, 7×7, and 9×9 for \agrmp, and the proportion of the extracted gradient to be top 10\%, 25\%, 50\%, and 75\% for \agrxai.
\subsubsection{Impact of Participant Number} \label{sec:partiNumber}
\begin{figure}
\vspace{-\baselineskip}
    \subfloat[500 Participants]{\includegraphics[height=1.1in]{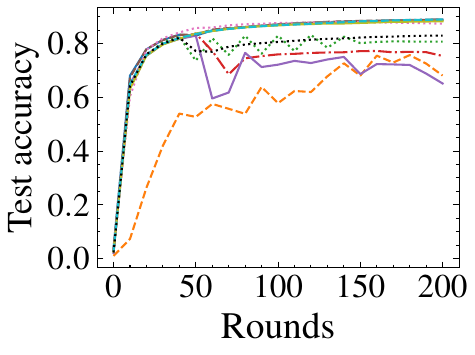}%
\label{fig:500}}
\subfloat[1000 Participants]{\includegraphics[height=1.1in]{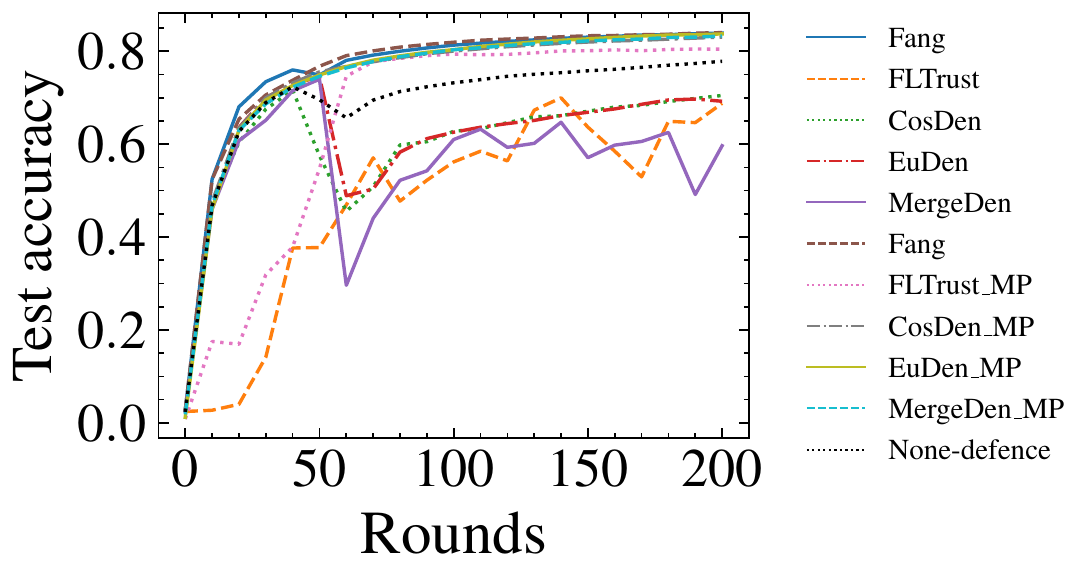}%
\label{fig:1000}}
\caption{Performance of run-time TA for untargeted attacks with a large number of participants.}
\label{fig:big}
\end{figure}

\begin{figure*}[!t]
\vspace{-\baselineskip}

 \subfloat[30-targeted \qquad \qquad 50-targeted \qquad \qquad 100-targeted]{\includegraphics[height=0.9in]{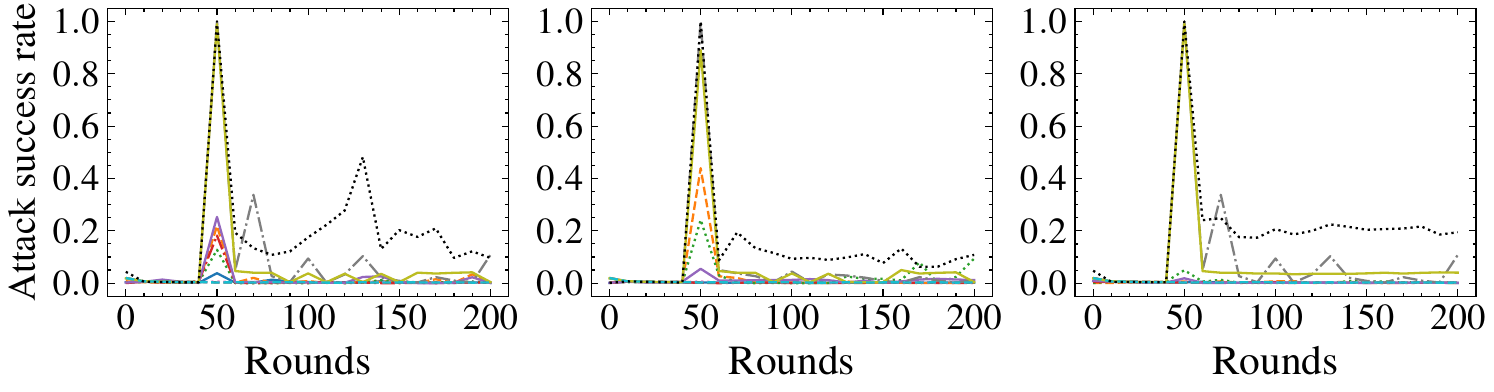}%
\label{fig:2a}}
\hfil
\subfloat[30-untargeted \qquad \qquad 50-untargeted \qquad \qquad 100-untargeted]{\includegraphics[height=0.9in]{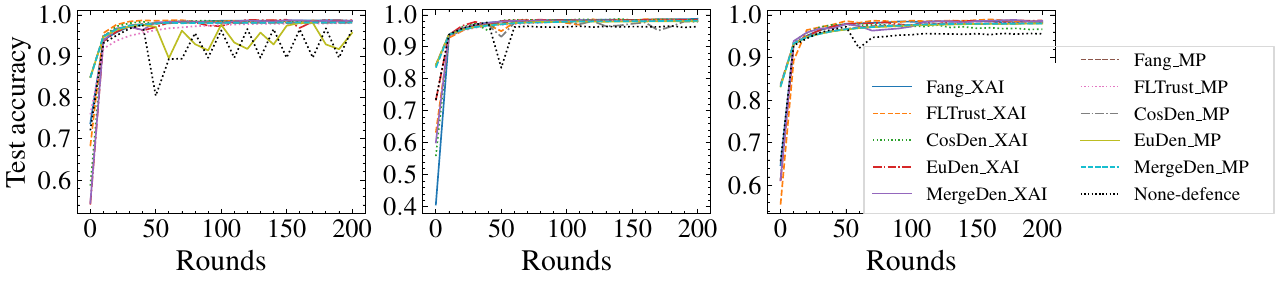}%
\label{fig:2b}}

\caption{(a) run-time ASR for targeted attacks, (b) run-time TA for untargeted attacks with different numbers of participants on MNIST.}

\label{fig:2}
\end{figure*}

Fig. \ref{fig:2} demonstrates the run-time ASR of \codename for targeted attacks and TA for untargeted attacks on MNIST, varying the participant count. Our findings reveal that \codename outperforms the base aggregator and achieves comparable performance to FedAvg when the number of participants ranges from 30 to 100.
Furthermore, we find that Base aggregators underperform when there is a large number of participants (500 and 1000 participants), leading to a reduction in overall performance for all considered aggregators. However, our proposed \agrmp still maintains good performance, as shown in Fig. \ref{fig:big}.

\subsubsection{Impact of Malicious Client Proportion} \label{sec:maliPortion}
\begin{figure*}[!t]
\centering
 \subfloat[0.2-targeted \qquad \qquad 0.3-targeted \qquad \qquad 0.4-targeted]{\includegraphics[height=0.9in]{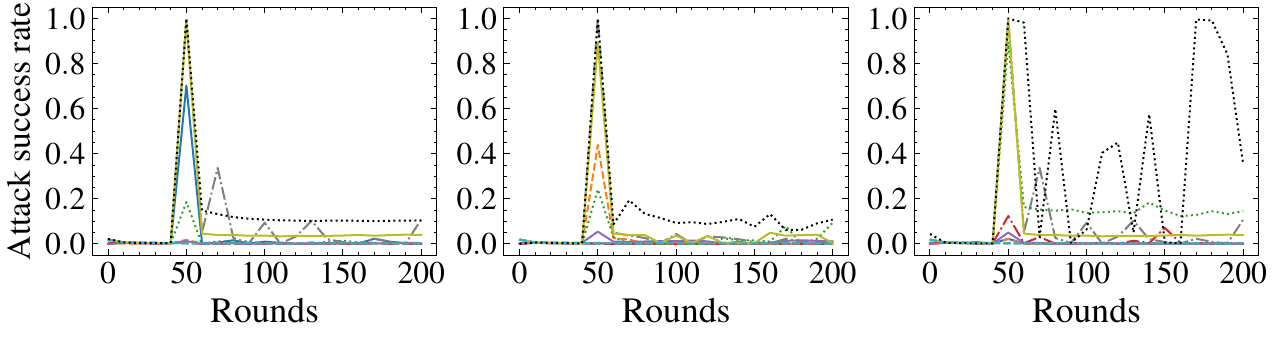}%
\label{fig:4a}}
\hfil
\subfloat[0.2-untargeted \qquad \qquad 0.3-untargeted \qquad \qquad 0.4-untargeted]{\includegraphics[height=0.9in]{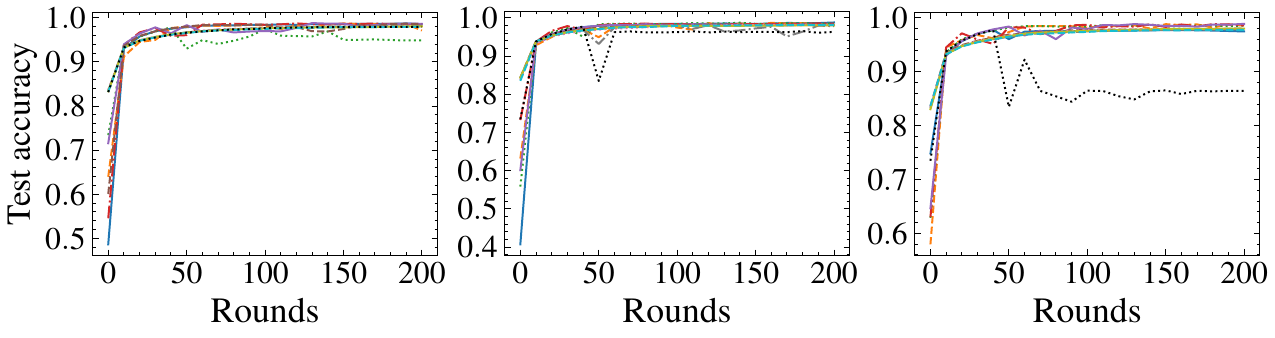}%
\label{fig:4b}}


\caption{(a) run-time ASR for targeted attacks, (b) run-time TA for untargeted attacks with different proportions of malicious participants on MNIST.}
\label{fig:4}
\end{figure*}

This section examines the effect of varying ratios of malicious participants ($M_f$) on the performance of \codename. The results are presented in Fig. \ref{fig:4}, which depicts the run-time TA and ASR of \codename on MNIST with $M_f$ values ranging from 0.2 to 0.4. The performance of FedAvg (without defense) is shown to drop significantly as the proportion of malicious participants increases. However, the performance of \codename remains stable and exhibits high TA even when $M_f$ reaches 40\%. 




\subsubsection{Impact of the Proportion of the Extracted Gradient}
\label{sec:top}
\begin{figure}[!t]
\vspace{-\baselineskip}
\centering

\subfloat[\cosmp]{\includegraphics[width=1.1in]{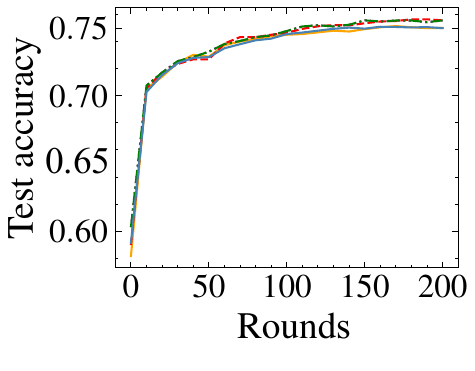}
\label{fig:6b}}
\hfil
\subfloat[\fangmp]{\includegraphics[width=1.1in]{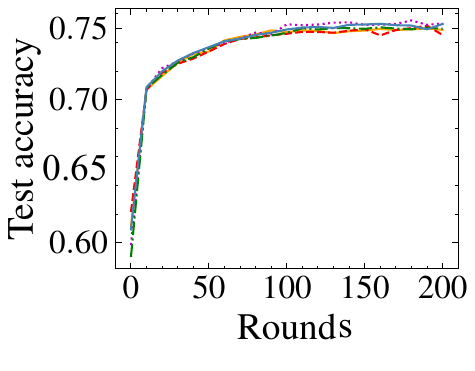}
\label{fig:6c}}
\hfil
\subfloat[\flmp]{\includegraphics[width=1.1in]{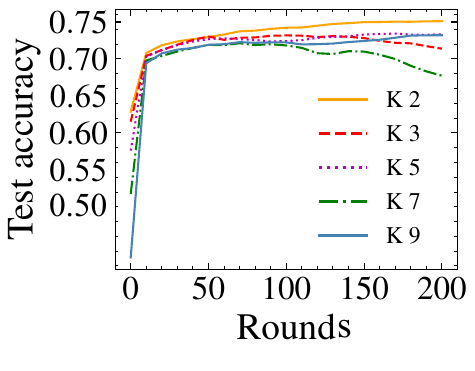}
\label{fig:6d}}
\caption{The impact of kernel size for \agrmp on CIFAR-10. }
\label{fig:kernel}
\end{figure}

In this section, we investigate how the proportion of the extracted gradient affects the performance of \codename. 
First, for \agrmp, the varying kernel sizes can affect the selection of the extracted gradients. Specifically, the kernel size is experimented within the range of 2×2, 3×3, 5×5, 7×7, and 9×9. The results indicate that the distance-based \cosmp and prediction-based \fangmp are not significantly impacted by the kernel size, whereas the trust bootstrapping-based \flmp shows a slight decrease in performance with larger kernel sizes such as 7×7 and 9×9. This mechanism computes the reference updates on a small trust set, which may result in non-negligible information loss when the kernel size is too large. The findings are presented in Fig. \ref{fig:kernel}.

\begin{figure}[!t]
\vspace{-\baselineskip}
\centering

\subfloat[\cosxai]{\includegraphics[width=1.1in]{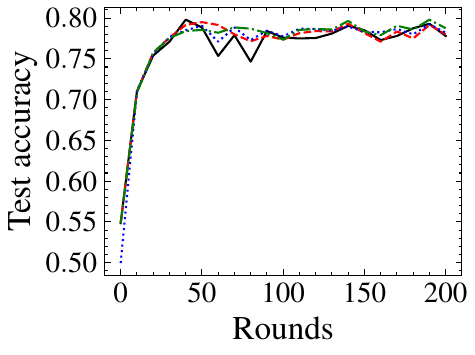}
\label{fig:6b}}
\hfil
\subfloat[\fangxai]{\includegraphics[width=1.1in]{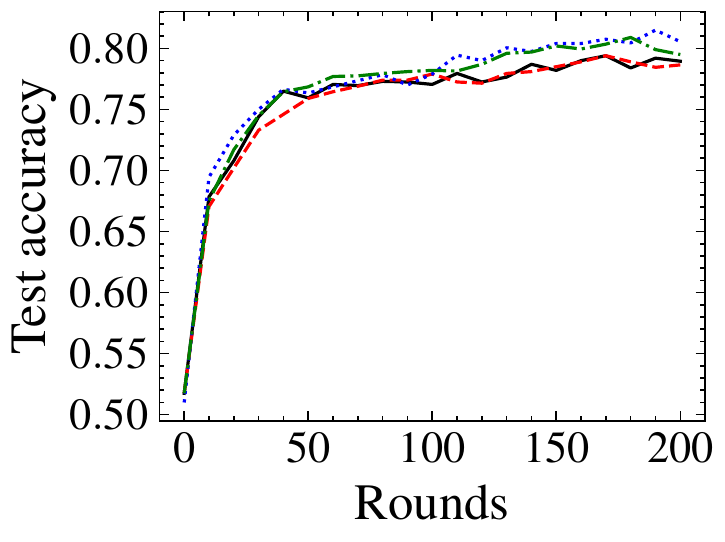}
\label{fig:6c}}
\hfil
\subfloat[\flxai]{\includegraphics[width=1.1in]{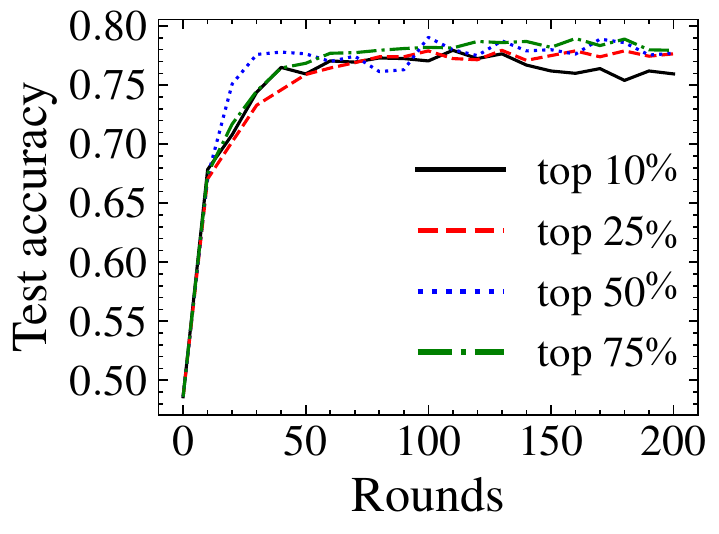}
\label{fig:6d}}
\caption{The impact of the proportion of the extracted gradient for \agrxai on CIFAR-10. }
\label{fig:top-three}
\end{figure}

\begin{figure}[!t]
\vspace{-\baselineskip}
\centering
\subfloat[Untargeted attacks]{\includegraphics[width=1.3in]{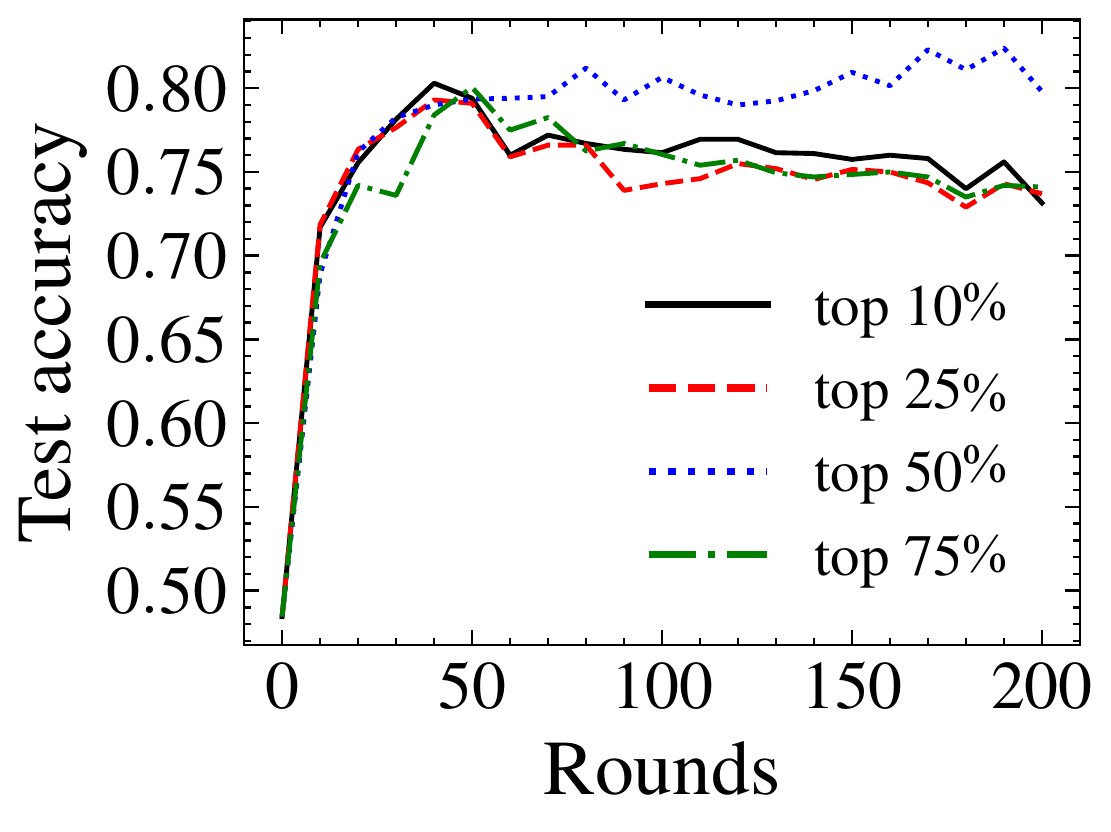}
\label{fig:7a}}
\hfil
\subfloat[Targeted attacks]{\includegraphics[width=1.3in]{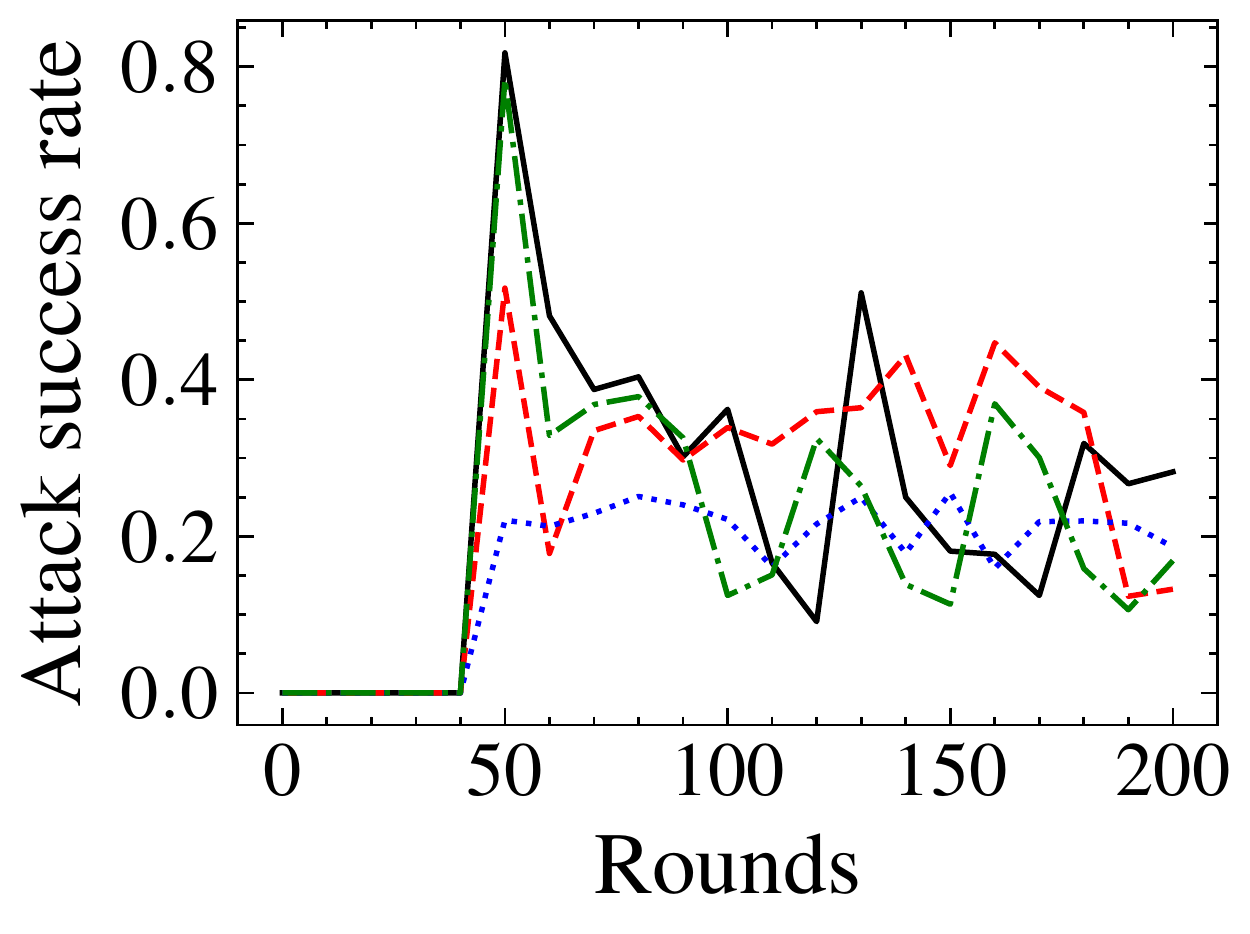}
\label{fig:7b}}

\caption{The impact of the proportion of the extracted gradient on CATvsDOG Kaggle. }
\label{fig:top}
\end{figure}

We also investigate how the proportion $p$ affects the performance of \agrxai. We conduct experiments by extracting the top 10\%, 25\%, 50\%, and 75\% of the gradient $g_i$. It shows a similar result as \agrmp, where the trust bootstrapping-based \flxai shows a slight decrease in performance with a large reduction size such as 10\%. The detailed findings are presented in Fig. \ref{fig:top-three} and \ref{fig:top}.
The results indicate that extracting the top 50\% of the gradients has the best performance for both untargeted and targeted attacks. This choice not only achieves the best evaluation metric but also exhibits the smoothest trend. 

\subsubsection{Impact of the validation datasets}
\label{sec:validation}
In the case of \fang and \fl-based defenses, which require the aggregator to hold a small validation dataset, we consider two scenarios \cite{cao_fltrust_2022} to evaluate the impact for validation dataset distribution:\\
\begin{itemize}
    \item Case I: In this scenario, we assume the aggregator has the capability to construct a well-balanced validation dataset that represents the distribution of the training data. To achieve this, we uniformly and randomly sample the validation dataset from the clean local training data of clients.\\
    \item Case II: We assume a validation dataset with a different distribution compared to the training data, specifically biased towards a particular class (class 1 in our case). We use bias probability ($\theta$) to represent the fraction of examples from that specific class and set the other $1-\theta$ examples to be uniformly sampled from the other classes.\\
\end{itemize}
Table \ref{table:bais} presents the model TA based on different bias probabilities ($\theta$), and the result shows that the impact of data distribution on these methods differs:\\
For the \fang-based methods, they consistently perform well across various bias probabilities.
In contrast, the \fl-based methods demonstrate varying performance. The base \fl method fails when the bias probability is 0.4. However, our \codename-equipped approach displays improved resilience and can handle a bias probability up to 0.9.
This difference in performance can be attributed to the underlying mechanisms of the base methods. The \fang-based approach evaluates local models based on their error rates on the clean validation dataset. It remains effective even in the presence of dataset bias, as the benign model continues to make accurate predictions.
On the other hand, the \fl-based methods train a clean model using its own clean dataset and assign trust scores to each update by comparing them with the clean model. Consequently, it becomes vulnerable to highly biased validation datasets. However, our approach mitigates this vulnerability by amplifying the differences between benign and malicious gradients. Even when trained on biased data, the gradients remain sufficiently close to the benign ones, facilitating the distinction of malicious gradients. This insight highlights the effectiveness of our approach in handling validation dataset biases.

\begin{table}[h]
\scriptsize
\centering
\caption{The TA of \fl and \fang based methods under \textbf{\texttt{\textbf{\texttt{G-asc}}}} attack on the CIFAR dataset when the validation dataset is sampled with different bias probabilities. We highlight 0.1 in bold, which indicates that the training fails to converge.} 
\label{table:bais}
\begin{tabular}{c|ccccccc}
\hline
Bias & 0    & 0.2  & 0.4          & 0.6  & 0.8          & 0.9          & 1            \\ \hline
None & 0.60 & 0.61 & 0.59         & 0.60 & 0.60         & 0.60         & 0.59         \\
T    & 0.57 & 0.57 & \textbf{0.1} & 0.55 & \textbf{0.1} & \textbf{0.1} & \textbf{0.1} \\
TM   & 0.59 & 0.58 & 0.58         & 0.57 & 0.57         & \textbf{0.1} & \textbf{0.1} \\
TX   & 0.58 & 0.58 & 0.58         & 0.57 & 0.58         & 0.58         & \textbf{0.1} \\
F    & 0.58 & 0.57 & 0.56         & 0.55 & 0.54         & 0.53         & 0.54         \\
FM   & 0.59 & 0.58 & 0.56         & 0.55 & 0.55         & 0.52         & 0.55         \\
FX   & 0.60 & 0.59 & 0.59         & 0.58 & 0.56         & 0.56         & 0.55         \\ \hline
\end{tabular}

\end{table}

%% file: chapters/discussion.tex
\section{Discussion and Limitations}
\label{sec:data-heter}
\subsection{Effect of Dataset's Heterogeneity}
Our observations indicate that the performance of the three categories of Byzantine-robust mechanisms varies depending on the dataset employed. Previous studies have shown that the generalization bound of ML is linked to the diversity of the training data \cite{arora_fine-grained_2019}. 
As the model's generalization capacity decreases, poisoning attacks become more effective, as it enhances the attacker's ability to increase the loss on poisoned examples. Conversely, a well-generalized model is better equipped to handle varied input and thus exhibits greater resilience to malicious injections.

Our experimental findings provide evidence of a correlation between the heterogeneity of datasets and the efficacy of defense mechanisms. We measure the heterogeneity of the datasets by computing the average intra-label cosine similarity using the approach outlined in \cite{arora_fine-grained_2019}. Specifically, given a training set label $\xi $ denoted as $D_\xi $, and the number of classes in the dataset denoted as $E$, the heterogeneity scores is computed as:
\begin{equation}
\label{eq:cos}
    1-\frac{1}{E} \sum_{\xi=1}^{E} \frac{\sum D_{\xi} \cdot D_{\xi}^{\top}}{\left\|D_{\xi}\right\|^{2}}.
\end{equation}

\begin{table}[]
\fontsize{20}{21}\selectfont
\hspace{-10pt}
\renewcommand{\arraystretch}{1.2}
\centering
\caption{Heterogeneity of the datasets.}
\label{table:heter}
\resizebox{0.5\textwidth}{!}{
\begin{tabular}{ccccccc}
\toprule
MNIST & CIFAR-10 & LOCATION30 & PURCHASE100 & TEXAS100 & CAT-DOG & FASHION-MNIST \\
0.15  & 0.24     & 0.12       & 0.03        & 0.61     & 0.41    & 0.21      \\
\bottomrule
\end{tabular}

}
\end{table}

Table \ref{table:heter} presents the heterogeneity scores for each dataset. Our analysis indicates that datasets with lower heterogeneity, as measured by larger intra-label cosine similarity, confer advantages to defenders, particularly against the scale attack. For instance, the scale attack on PURCHASE100 and LOCATION30 exhibits the lowest success rates (less than 10\%) when defensive mechanisms are applied. Furthermore, we find that the performance of trust bootstrapping-based mechanisms is significantly impacted when the training data's heterogeneity is extremely high, as observed in TEXAS100, with the highest heterogeneity score.

\subsection{Limitations Inherited from the Base Mechanisms}
It has been recognized that the \codename approach may not be able to overcome certain inherent constraints of the underlying mechanisms. For instance, in the case of trust bootstrapping-based mechanisms, \codename may prove ineffective if the trusted set is poisoned \cite{cao_fltrust_2022}. Additionally, for certain base aggregation rules, such as distance-based aggregators, which necessitate knowledge of the malicious fraction $M_f$, \codename also relies on this assumption. However, it may not always be feasible for the aggregator to obtain such information in practical applications.

\subsection{Comparison of \agrmp and \agrxai}
\label{sec:compare}
In terms of robustness, \agrxai outperforms \agrmp, especially for targeted attacks. The enhanced performance of \agrxai can be attributed to its ability to capture significant features containing triggers, thereby improving detection performance. In terms of fidelity, \agrxai also outperforms \agrmp in general, but both \agrmp and \agrxai show significantly less fidelity loss compared to the base \agrs. 
In terms of efficiency, \codename reduces gradient size, resulting in reduced time consumption, particularly for larger neural network architectures. However, the \agrxai approach demonstrates higher time consumption due to additional computations required for individual update feature weights. 

Specifically, \exai demonstrates the highest robustness against targeted attacks, while \fangxai performs the best against untargeted attacks. In terms of fidelity, \cosmp exhibits superior performance, closely followed by \cosxai. Furthermore, \flmp demonstrates superior efficiency with small to medium network sizes, while \cosmp delivers the best efficiency with large networks.

In conclusion, investigating the trade-off between robustness, fidelity, and efficiency in the ten proposed versions is a promising direction for future research. By considering the factors mentioned above and conducting rigorous experiments, researchers can gain valuable insights into the interplay between these aspects and make informed decisions to achieve the desired balance in real-world scenarios.

\subsection{Furture Works}
Several promising directions for future research are of interest. Firstly, we aim to explore and establish a better trade-off between robustness, fidelity, and efficiency. Secondly, ongoing efforts are dedicated to enhancing the explainability of Natural language processing (NLP) tasks~\cite{nlp, sisodiaexplainable}. Therefore, our \agrxai-based approach, which leverages Explainable AI (XAI), holds promise for extension into NLP tasks and improves the identification and understanding of significant features within participants' updates. We will also investigate other defense mechanisms, such as frequency-domain analysis for detecting malicious updates and style transformation techniques for converting malicious updates into benign ones.

%% file: chapters/conclusion.tex
\section{Conclusion}
This study presents \codename, a mechanism specifically designed to enhance the robustness, fidelity, and efficiency of FL methods against Byzantine adversaries. 
Through comprehensive evaluations across various scenarios, our approach demonstrates an average reduction of 40.07\% in the ASR while maintaining high levels of fidelity and efficiency. These results highlight the effectiveness of \codename in mitigating the impact of Byzantine adversaries in FL.
Furthermore, \codename presents the first work that incorporates XAI into the domain of malicious detection. This integration not only enhances the performance of existing methods but also opens up new avenues for future research and exploration. 


%% file: chapters/Appendix.tex


\newpage
\section*{Supplemental Document for \codename}

\subsection{Algorithm: \codename for Distance-based Mechanism}
\label{sec:a1}
The proposed algorithm involves processing the original gradients through an amplifier (line 1 in Algorithm \ref{alg:alg-distance}), followed by a distance measurement step that employs a pairwise comparison approach, such as Cosine similarity or Euclidean distance. This distance measurement step determines the density of each gradient's neighborhood, consisting of $K$ neighbors, where $K$ is set to a value greater than $N/2$, with $N$ representing the number of participants. The algorithm considers gradients residing in denser neighborhoods as benign ones. For each participant's update, the algorithm selects the K-nearest neighbors based on their similarity score and calculates the cumulative score in their neighborhood (lines 4-10). The top $N_t$ participants with higher scores are then added to the white list, which can be considered a hyperparameter chosen by the practitioner and submitted for aggregation (lines 12-19). Therefore, the proposed algorithm utilizes a novel approach to improve the performance of FL by selecting benign participants through a similarity-based analysis of gradient neighborhoods.
\begin{algorithm}
\renewcommand{\algorithmicrequire}{\textbf{Input:}}
\renewcommand{\algorithmicensure}{\textbf{Output:}}
\renewcommand{\algorithmicprocedure}{\textbf{function}}
\begin{algorithmic}[1]
\caption{\codename for Distance-based Mechanism.}\label{alg:alg-distance}
\Require {$g_i,g_j$ - the gradients collected from $i$-th and $j$-th clients; 
$g_{amp}^i$ - the amplified gradients from client $i$;
$N$ - number of participants;
$M_f$ - the fraction of malicious participants;
$K$ - number of neighbors examined.}
\Ensure $G_{detox}$ the detoxed gradients collection.
\State $\left\{g_{a m p}^{1}, g_{a m p}^{2}, \ldots\right\} \leftarrow \text { \codename }\left(g_{1}, g_{2}, \ldots\right)$
\For{\texttt{$i = 0,1,2,...,N$}}
        \For{\texttt{$j = 0,1,2,...,N$}}
        \State // Pair-wise cosine similarity
        \State $C_{i,j} \leftarrow \frac{g_{a m p}^{i} \cdot g_{a m p}^{j}}{\left\|g_{a m p}^{i}\right\| \cdot\left\|g_{a m p}^{j}\right\|}$    
        \EndFor
        \State // Sum up the similarity of the $K$-nearest neighborhood as $S_i$
        \State $C_{i} \leftarrow \text { Descending-sort }\left(C_{i,j} \mid j=1,2,3, \ldots N\right)$
        \State $S_{i} \leftarrow \sum (C_{i,j} \mid j=1,2,3, \ldots K)$
    \EndFor
\State $\text { Whitelist } \leftarrow \emptyset$
\For{\texttt{$i = 0,1,2,...,N$}}
\State // Add into white-list of $i$ is with larger $S_i$
\If{$S_i$ in the largest $(1-M_f)*N$ values $\forall S_i$} 
                \State $\text { Whitelist } \leftarrow \text { Whitelist } \cup\{i\}$  
            \EndIf 
\EndFor
\State $G_{\text {detox }} \leftarrow\left\{g_{i} \mid i \in \text { Whitelist }\right\}$
    \State \Return $G_{detox}$

\end{algorithmic}

\end{algorithm}

\subsection{Algorithm: \codename for Prediction-based Mechanism}
\label{sec:a2}
The \codename algorithm initially performs gradient amplification while ensuring that the size of the collected gradients is restored (line 1 in Algorithm \ref{alg:alg-prediction}). This is accomplished by retaining the maximum value in each feature map and replacing the dropped features with zeros. Subsequently, the restored gradients undergo calculation of LRR and ERR \cite{fang_local_2021}, and those leading to superior prediction performance are included in the white list (line 3).
\begin{algorithm}
\renewcommand{\algorithmicrequire}{\textbf{Input:}}
\renewcommand{\algorithmicensure}{\textbf{Output:}}
\renewcommand{\algorithmicprocedure}{\textbf{function}}
\begin{algorithmic}[1]
\caption{\codename for Prediction-based Mechanism.}\label{alg:alg-prediction}
\Require {
$G$ - the collected gradients, equivalent to $\left\{g_{1}, g_{2}, g_{3}, \ldots\right\}$; 
$G_{amp}$ - the amplified gradients; 
$g_i$ - the collected gradient from $i$-th client in $G$.}
\Ensure $G_{detox}$ the detoxed gradients collection.
\State $G_{a m p} \leftarrow \operatorname{\codename}(G , Restore-size  =  True  )$
\State Whitelist  $\leftarrow \operatorname{LRR}\left(G_{a m p}\right) \cap \operatorname{ERR}\left(G_{a m p}\right) $
\State $G_{\text {detox }} \leftarrow\left\{g_{i} \mid i \in\right.$  Whitelist, $ \left.g_{i} \in G\right\}$
\State \Return $G_{detox}$
\end{algorithmic}

\end{algorithm}
\subsection{Algorithm: \codename for Trust Bootstrapping-based Mechanism}
\label{sec:a3}
The \codename algorithm not only conducts gradient amplification on the collected gradients to produce $g_{amp}^i$ (line 1 in Algorithm \ref{alg:alg-trust}) but also amplifies the gradient generated from the trusted validation set for trust bootstrapping to produce $g_{amp}^0$ (line 2). The trust score $TS_i$ for each participant is subsequently determined by computing the ReLU-clipped cosine similarity between the amplified gradients $g_{amp}^i$ and $g_{amp}^0$ (line 4). The collected local model update magnitudes are then normalized (line 5), weighted by their respective trust scores, and aggregated to form the global model (line 7).
\begin{algorithm}
\renewcommand{\algorithmicrequire}{\textbf{Input:}}
\renewcommand{\algorithmicensure}{\textbf{Output:}}
\renewcommand{\algorithmicprocedure}{\textbf{function}}
\begin{algorithmic}[1]
\caption{\codename for Trust Bootstrapping-based Mechanism.}\label{alg:alg-trust}
\Require {
$g_{amp}^i$ - the amplified gradients from client $i$;
$g_i$ - the collected gradient from client $i$;
$g_0$ - the gradient generated from the trusted root dataset.}
\Ensure $G_{detox}$ the detoxed gradients collection.
\State $\left\{g_{a m p}^{1}, g_{a m p}^{2}, \ldots\right\} \leftarrow \text { \codename }\left(g_{1}, g_{2}, \ldots\right)$
\State $g_{a m p}^{0} \leftarrow \text { \codename }\left(g_{0}\right)$
\For{$i=1,2, \ldots, N$}
    \State $\qquad T S_{i} \leftarrow \mathbf{R e L U}\left(\frac{g_{a m p}^{i} \cdot g_{a m p}^{0}}{\left\|g_{a m p}^{i}\right\| \cdot\left\|g_{a m p}^{0}\right\|}\right)$
    \State $\qquad \bar{g}_{i} \leftarrow \frac{\left\|g_{0}\right\|}{\left\|g_{i}\right\|} \cdot g_{i}$
\EndFor
\State $g_{\text {detox }} \leftarrow \frac{1}{\sum_{j=1}^{N} T S_{j}} \sum_{i=1}^{N} T S_{i} \cdot \bar{g}_{i}$
\State \Return $G_{detox}$
\end{algorithmic}

\end{algorithm}